\documentclass[a4paper,11pt]{article}
\pdfoutput=1
\usepackage{jheppub}
\usepackage[T1]{fontenc}
\usepackage{subfigure}
\usepackage{overpic}
\usepackage{multirow}
\usepackage{lineno}

\newcommand{\ksks}{K^0_SK^0_S}

\newcommand{\lum}{{\cal L}}
\newcommand{\pip}{\pi^+}
\newcommand{\pim}{\pi^-}

\newcommand{\etac}{\eta_c}
\newcommand{\hc}{h_{c}}
\newcommand{\psip}{\psi'}
\newcommand{\jpsi}{J/\psi}
\newcommand{\ks}{K_{S}^{0}}
\newcommand{\EE}{e^+e^-}
\newcommand{\pp}{\pi^+\pi^-}
\newcommand{\kk}{K^+K^-}
\newcommand{\DD}{D\bar{D}}

\newcommand{\mev}{\rm{MeV}}
\newcommand{\mevcc}{{\rm MeV}/c^2}
\newcommand{\gev}{\rm{GeV}}
\newcommand{\gevcc}{{\rm GeV}/c^2}

\title{\boldmath Search for $\EE\to \ks\ks\hc$}

\collaborationImg{\includegraphics[height=3cm,angle=90]{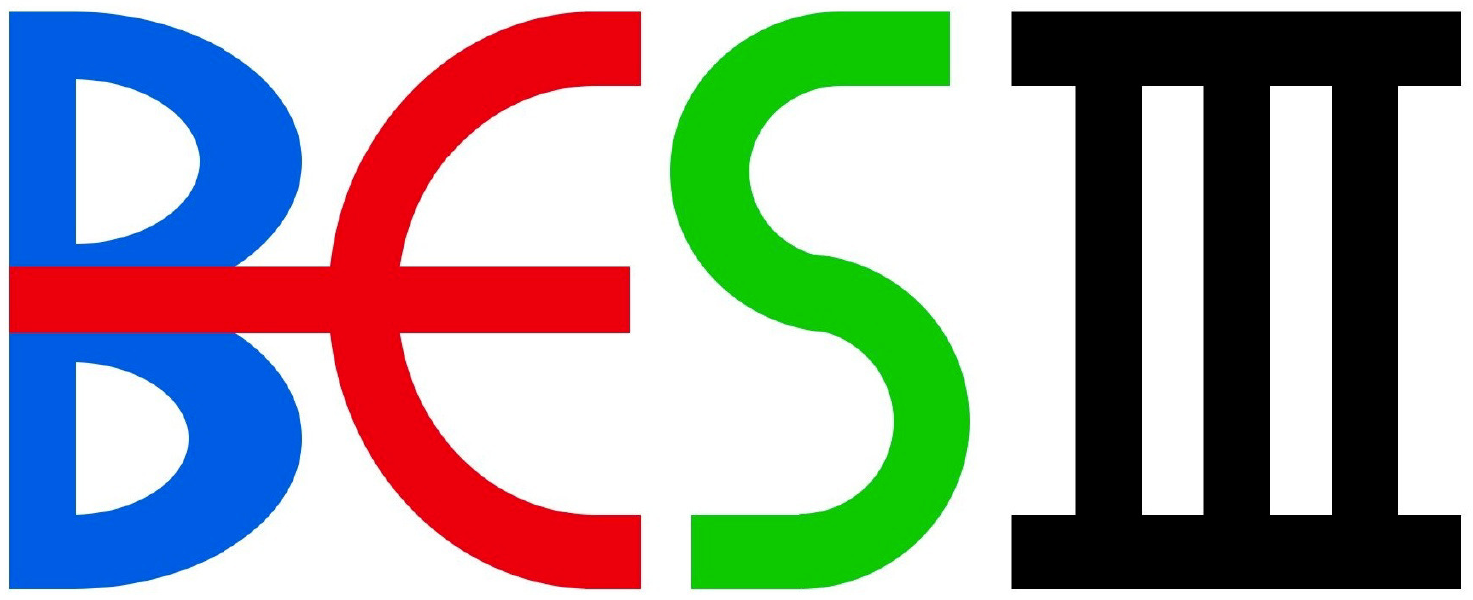}}
\collaboration{The BESIII Collaboration}

\keywords{$\EE$ collision, Exotic hadron, Quarkonium}

\emailAdd{besiii-publications@ihep.ac.cn}

\arxivnumber{....}

\abstract{
Using $e^+e^-$ collision data at 13 center-of-mass 
energies ranging from $4.600$ 
to $4.951~\rm{GeV}$ collected with the BESIII detector,
we conduct the first search for the $\EE\to\ksks\hc$ process and investigate the resonance structures in the cross section line shape.
No significant signal 
is observed, and the upper limits of the Born cross sections at each center-of-mass energy
are presented.
The ratio $\frac{\sigma(\EE\to\ks\ks h_c)}{\sigma(\EE\to\ks\ks\jpsi)}$
is determined to be $0.15 \pm 0.22$.
This result indicates that if vector 
states exist in this energy region,
their decay into $\hc$ is significantly 
suppressed compared to decays into $\jpsi$.

}

\begin{document} \maketitle \flushbottom

\section{Introduction}

Over the last two decades, the discovery of 
numerous charmonium-like states has 
significantly expanded our understanding of charmonium spectroscopy.
In particular, a series of vector states ($J^{PC} = 1^{--}$),
such as the $Y(4260)$ and 
$Y(4660)$~\cite{intro-BaBar-pipijpsi,PDG,intro-Belle-pipipsip},
have been discovered above the $\DD$ threshold.
These $Y$ states cannot be easily classified as conventional charmonium 
states and are thus considered potential candidates for hadrons with exotic 
internal structures. Hypotheses for their composition include hybrid~\cite{Zhu:2005hp, Close:2005iz, Kou:2005gt},
tetraquark~\cite{Maiani:2014aja}, 
molecule~\cite{Cleven:2013mka, Ding:2008gr, Wang:2013cya},
hadrocharmonium states~\cite{Dubynskiy:2008mq, Li:2013ssa},
or kinematically induced peaks~\cite{Chen:2017uof}.

Recently, the BESIII collaboration studied the processes
$\EE\to\kk\jpsi$~\cite{BESIII:2022joj,BESIII:2023wqy}
and $\EE\to\ksks\jpsi$~\cite{intro-BESIII-ksksjpsi}.
The cross section line shapes revealed two structures,
referred to as $Y(4500)$ and $Y(4710)$, 
around $4.5$ and $4.7$~GeV, respectively.
Furthermore, in the process $\EE\to D_s^{*+}D_s^{*-}$,
a structure near $4.75$~GeV was needed to describe
the cross section line shape~\cite{BESIII:DssDss}.
These structures are observed in final states containing $s\bar{s}c\bar{c}$
quarks.
Studying these $Y$ states across various processes is crucial for 
advancing our understanding of their nature.
The decay of conventional vector charmonium states into $h_c$ is expected 
to be suppressed due to heavy-quark spin symmetry~\cite{Oncala:2017hop};
thus, searches for $Y$ states decaying into $\hc$ could provide valuable insight into their exotic properties.

To date, the process $\EE\to K \bar{K} \hc$ remains unobserved. This motivated
us to  try to measure the cross section line shape of $\EE\to K \bar{K} \hc$
and to search for potential $Y$ states.
In this analysis, we present a study of the process $\EE\to\ksks\hc$
using 13 data samples collected within the energy range from $4.600$ to
$4.951~\gev$ with the BESIII detector~\cite{Ablikim:2009aa} at the BEPCII
collider~\cite{Yu:IPAC2016-TUYA01}.
The $h_c$ meson is reconstructed via its radiative decay to $\eta_c$.
Omitting the reconstruction of the latter, we employ
a partial reconstruction method to improve the detection statistics for the signal process.

\section{BESIII detector and data samples}

The BESIII detector~\cite{Ablikim:2009aa} records $e^+e^-$ collisions
provided by the BEPCII storage ring~\cite{Yu:IPAC2016-TUYA01} in the
center-of-mass (c.m.) energy range from 1.84 to 4.95~GeV, with a peak  
luminosity of $1.1\times10^{33}$~cm$^{-2}$s$^{-1}$ achieved at
$\sqrt{s}=3.773~\gev$.
The cylindrical core of the BESIII detector covers 93\% of the full solid angle and consists of a helium-based
multilayer drift chamber~(MDC), a plastic scintillator time-of-flight
system~(TOF), and a CsI(Tl) electromagnetic calorimeter~(EMC),
which are all enclosed in a superconducting solenoidal magnet
providing a 1.0~T magnetic field. The solenoid is supported by an
octagonal flux-return yoke with resistive plate counter muon
identification modules interleaved with steel.
The charged-particle momentum resolution at $1~{\rm GeV}/c$ is
$0.5\%$, and the $dE/dx$ resolution is $6\%$ for electrons
from Bhabha scattering. The EMC measures photon energies with a
resolution of $2.5\%$ ($5\%$) at $1$~GeV in the barrel (end cap)
region. The time resolution in the TOF barrel region is 68~ps, while
that in the end cap region is 110~ps. The end cap TOF
system was upgraded in 2015 using multi-gap resistive plate chamber
technology, providing a time resolution of 60~ps~\cite{etof}.
All the data samples, except for the one at 4.6 GeV,
benefit from the TOF upgrade.

The data samples used for this analysis were collected 
at 13 c.m. energies ($\sqrt{s}$) ranging from $4.600$ to $4.951~\gev$.
The c.m. energies are measured by selecting di-muon
or $e^+e^-\to\Lambda_c\bar{\Lambda}_c$
events~\cite{cms-offline-XYZ, cms-lumi-round1314}
with an uncertainty of 0.6$~\mev$.
The total integrated luminosity is $6.4$~fb${}^{-1}$
with an uncertainty of 1.0\%,
determined by selecting large angle
Bhabha scattering events~\cite{BESIII-lumi-yifan, cms-lumi-round1314}.
The process $\EE\to\ksks\jpsi$ is used as a control sample
to determine the mass resolution difference between data and simulation.
Considering the momentum range of $\ks$,
data samples with $4.19<\sqrt{s}<4.29$~GeV,
corresponding to an integrated luminosity of $9.5~\rm{fb}^{-1}$,
are used to select the control sample events.

Simulated samples are produced with 
a {\sc geant4}-based~\cite{geant4} Monte Carlo (MC) package, 
which includes the geometric description of
the BESIII detector and the detector response.
The simulation models the beam
energy spread and initial state radiation (ISR) in the $e^+e^-$
annihilations with the generator {\sc kkmc}~\cite{ref:kkmc}.
All particle decays are modeled with {\sc evtgen}~\cite{ref:evtgen}
using branching fractions either taken from the Particle Data
Group (PDG)~\cite{PDG}, when available,
or otherwise estimated with {\sc lundcharm}~\cite{ref:lundcharm}.
Final state radiation from charged final state particles 
is incorporated using {\sc photos}~\cite{photos}.

Inclusive MC samples, which include the production of open-charm mesons,
the ISR production of vector charmonium(-like) states,
and continuum processes, are used to study the background contributions.
A signal MC sample of $\EE\to\ks\ks\hc$ with $\hc$ and $\etac$ decaying
inclusively, is used to determine the detection efficiencies.
For the non-resonant three-body signal process 
$\EE\to \ksks\hc$,
the momenta distributions of final state particles
are generated following phase space.
The cross section of $\EE\to\ksks\hc$ is assumed to follow the three-body decay phase space factor.
For the dominant decay of $\hc$, $\hc\to\gamma\etac$,
the angular distribution of the $E1$ photon (in the $\hc$ rest frame)
is generated as $1+\cos^2\theta$.
A MC sample of $\EE\to\ksks\jpsi$ is simulated with
final state particles generated following phase space.
The cross section line shape from a previous
measurement~\cite{intro-BESIII-ksksjpsi} 
is used as input in the simulation of $\EE\to\ksks\jpsi$ events.

\section{Event selection}

In this analysis, the signal process $\ks\ks\hc$
is reconstructed by selecting two $\ks$ mesons
and one photon for the partial reconstruction of the
$\hc\to\gamma\etac$ decay.

\begin{figure*}[htbp]
\begin{center}
\includegraphics[width=0.48\textwidth]{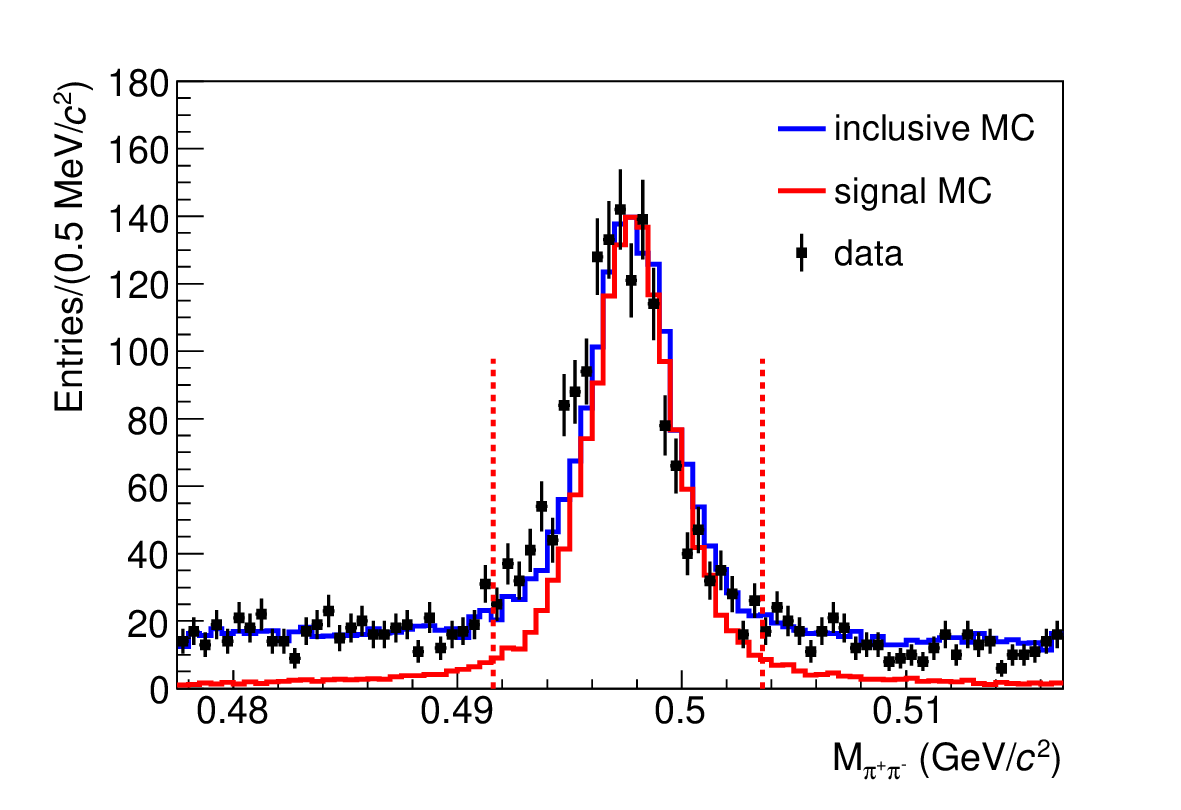}
\includegraphics[width=0.48\textwidth]{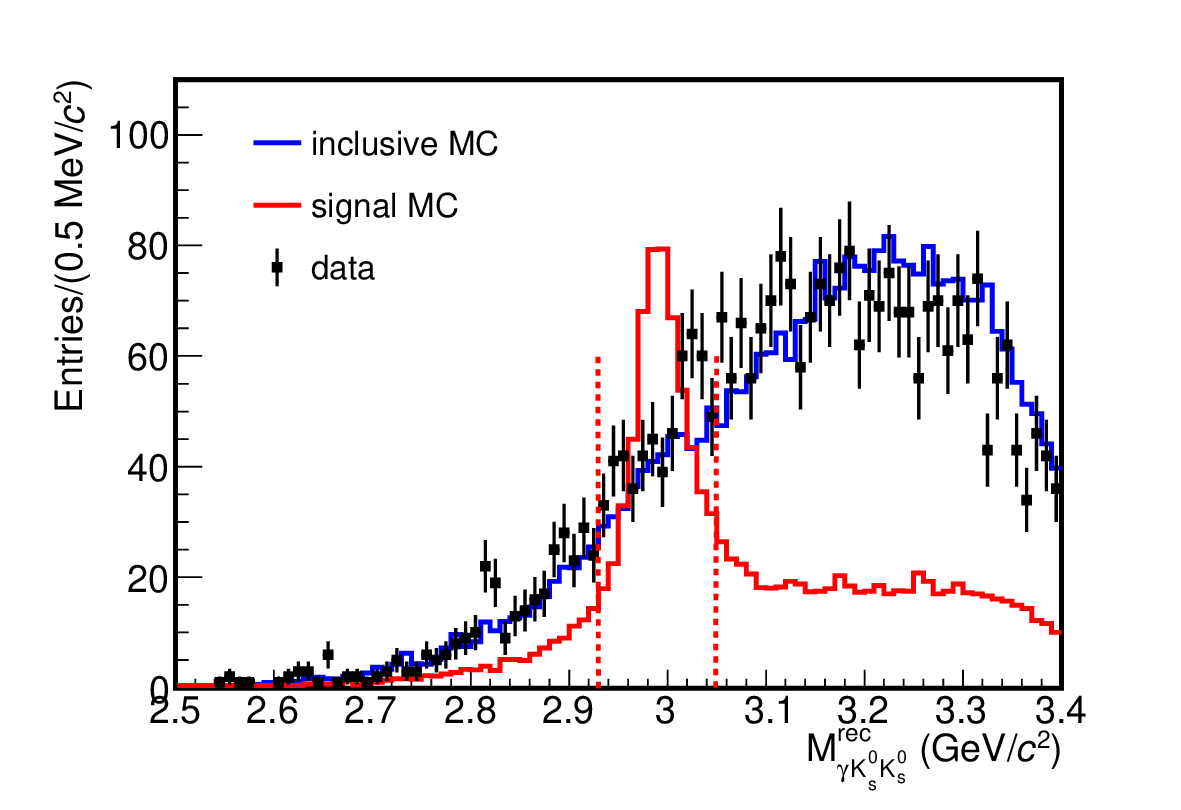}
\end{center}
\caption{
Distributions of $M_{\pp}$ (left)
and $M_{\gamma\ksks}^{\rm{rec}}$ (right).
The black dots with error bars are from data,
the blue curves are inclusive MC,
the red curves are signal MC,
and the vertical dashed lines indicate the optimized selection criteria.
$\hc$ decays inclusively in the signal MC sample.
Distributions from inclusive MC samples are normalized
according to the integral luminosity.
Distributions from signal MC samples are normalized
according to maximum.
}
\label{fig:opti ks}
\end{figure*}

A $\ks$ candidate is reconstructed from two oppositely charged tracks,
which are assigned as $\pip\pim$ without imposing 
further particle identification criteria.
These tracks are constrained to originate from a common vertex
and are required to have an invariant mass ($M_{\pp}$) within 
$(m_{\ks}\pm 6)~\mevcc$, where $m_{\ks}$ is the nominal mass of $\ks$ from
PDG~\cite{PDG}.
The size of the mass window is determined by performing 
an optimization based on the Punzi Figure-of-Merit (FoM)
defined as $\rm{FoM} = \frac{\epsilon}{A/2+\sqrt{B}}$,
where $A$ is set to be 3 as the expected significance,
$\epsilon$ is the efficiency given by the signal MC sample,
and $B$ represents the number of background events estimated by
the inclusive MC sample normalized according to the integrated luminosity.
The $M_{\pp}$ distribution and the corresponding mass window are
shown in Figure~\ref{fig:opti ks} (left).
Additionally, the decay length of the $\ks$ candidate is required
to be greater than twice the vertex resolution away from
the interaction point (IP).

A photon candidate is identified using showers in the EMC. 
The deposited energy of a shower must be more than
$25$~MeV in the barrel region,
where the polar angle $\theta$ satisfies $|\cos\theta|<0.8$,
and more than $50$~MeV in the end cap region
($0.86<|\cos\theta|<0.92$).
To suppress electronic noise and showers unrelated to the event, 
the difference between the EMC time and the event start time
is required to be within [0, 700]~ns.
Each signal candidate event is required to contain at least one photon.

After the selections described above, each signal candidate event must 
contain at least one $\ksks$ pair with 
each pion is used only once.
If multiple $\ksks$ pairs exist in an event, 
all combinations are retained for further analysis.
Each $\ksks$ combination is then paired with the photons in the event.
The photon with the recoil mass of $\gamma\ksks$
($M_{\gamma\ksks}^{\rm{rec}}$) closest to the nominal $\etac$ mass
($m_{\etac}$) is tagged as the $E1$ photon.
The $\gamma\ksks$ combinations are required to satisfy
$M_{\gamma\ksks}^{\rm{rec}}\in[2.94, 3.06]~\gevcc$,
a range optimized based on the Punzi FoM.
The $M_{\gamma\ksks}^{\rm{rec}}$ distribution and the mass window 
are shown in Figure~\ref{fig:opti ks} (right).

The background contribution from multi $\ks\ks$ combinations in the signal 
process is studied using signal MC simulations with a 
match method.
This method compares the 3-momentum of the reconstructed
charged pion tracks with the generator information.
An observable, defined as 
$\chi^2_{\rm{match}} = \Sigma_{i=1}^{4}(
\vec{p}_{\rm{rec.}, i} - \vec{p}_{\rm{gen.}, i})^2$,
is employed to indicate the goodness of the generator match,
where $\vec{p}_{\rm{rec.}, i}$
represents the 3-momentum of the $i$-th pion from the $\ks$ decay
after reconstruction, and $\vec{p}_{\rm{gen.}, i}$
denotes the 3-momentum of the $i$-th pion from the $\ks$ decay
at the generator level.
If there is more than one such combination in an event, 
the one with smallest $\chi^2_{\rm match}$ is selected. 
The combinations satisfying $\chi^2_{\rm{match}}<0.05$ 
are classified as matched.
The remaining combinations are identified as 
combinatorial backgrounds from the signal process.
This background contribution is
small and smoothly distributed across
the $\ksks$ recoil mass ($M_{\ksks}^{\rm{rec}}$) spectrum.

The same selection criteria are applied to the inclusive MC sample
to investigate background contributions from other processes.
The background events are found to be dominantly originating from processes 
with multiple light hadrons in the final state and smoothly distributed in 
the $M_{\ksks}^{\rm{rec}}$ distribution.
The $M_{\ksks}^{\rm{rec}}$ distribution from the 
inclusive MC sample, normalized according to integrated luminosity, is shown 
in Figure~\ref{fig:kskshc fit} as the brown histogram.

\section{Cross section}\label{chap:fit}

A fit to the $M_{\ksks}^{\rm rec}$ distribution is
performed using an unbinned maximum likelihood method
to determine the number of $\EE\to\ksks\hc$ signal events.
The signal contribution is described by the signal MC shape, 
convoluted with a Gaussian function to account for 
the resolution differences between the MC sample and the data.
The signal MC shape is parameterized,
with parameters fixed to those determined from the signal MC sample.
The parameters of the convoluted Gaussian function,
$\delta_{\rm mean}$ and $\delta_{\sigma}$, 
are derived from the control sample $\EE\to\ksks\jpsi$~\cite{intro-BESIII-ksksjpsi}.
The selection criteria for the $\ksks$ pair in the control sample
are identical to those used in the signal process.
The resolution differences are determined to
be $\delta_{\rm{mean}} = (0.19 \pm 0.55)~\mev$
and $\delta_{\rm{\sigma}} = (-0.01 \pm 1.07)~\mev$.

\begin{figure*}[htbp]
\begin{center}
\includegraphics[width=0.95\textwidth]{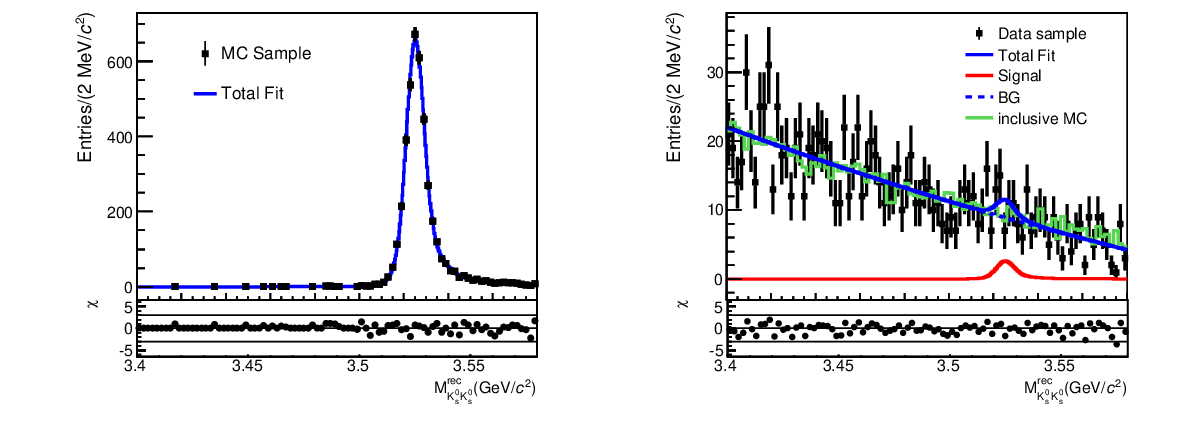}
\end{center}
\caption{The fit to the $M_{\ksks}^{\rm rec}$ distribution
from $\EE\to\ksks\hc$ process
at $\sqrt{s}=4.750$~GeV.
The left panel shows the fit to signal MC,
where the black dots with error bars and
blue solid line represent the 
signal MC and total fit curve, respectively.
The right panel shows the fit to data,
where the blue solid line is the total fit curve,
the blue dashed line is the background contribution,
and the red line is the fitted signal contribution.
The contribution from the inclusive MC sample is shown
as the brown histogram.
The $\chi$ distributions are shown below.
}
\label{fig:kskshc fit}
\end{figure*}

The background contribution for the $\EE\to\ksks\hc$ process
is described with a second-order Chebyshev function
with the parameters kept float.
The fit results for the $M_{\ksks}^{\rm rec}$ distribution from the signal MC and data at 
$\sqrt{s}=4.750$~GeV are shown in Figure~\ref{fig:kskshc fit}.
The bottom panels display the distributions of $\chi=\frac{N_{\rm data}-N_{\rm fit}}{\delta_{\rm data}}$,
where $N_{\rm data}$, $N_{\rm fit}$, and $\delta_{\rm data}$
represent the number of events from the data sample,
the total fit curve, and the statistical uncertainty of the data, respectively.
Fit results for the other data samples are presented in the appendix.

\begin{figure*}[htbp]
\begin{center}
\includegraphics[width=0.5\textwidth]{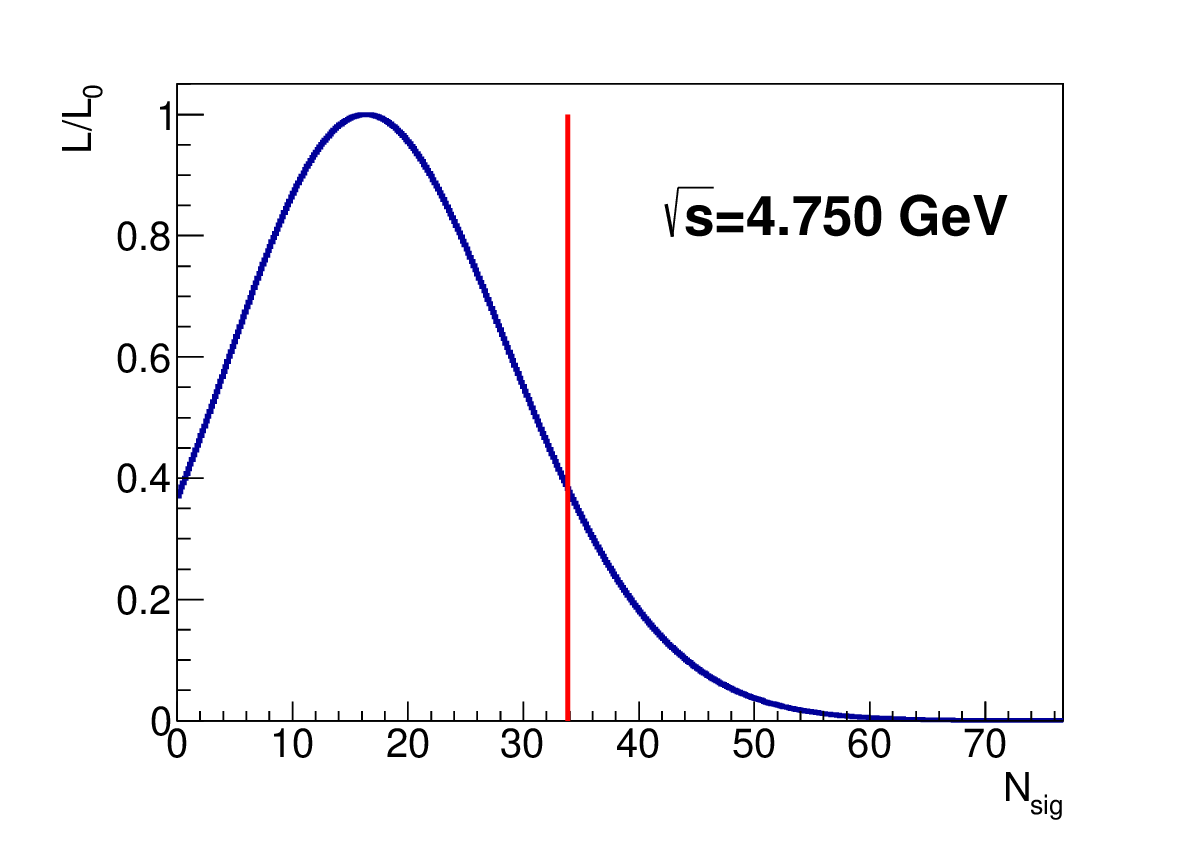}
\end{center}
\caption{The likelihood scan result
at the $\sqrt{s}=4.750$~GeV data sample.
The red vertical line indicates the position 
of the U.L. at the 90\% C.L.
}
\label{fig:ksksJpsi scan}
\end{figure*}

The signal significance for each data sample is evaluated by comparing
the difference of $(-\ln L)$ with and without the signal component,
where $L$ denotes the likelihood value.
The significance is found to be less than $2\sigma$ for each data sample,
and is not calculated if the nominal signal yield is negative.
The upper limit (U.L.) on the number of signal events is
determined at the 90\% confident level (C.L.) through a likelihood scan.
The distribution of $L/L_0$ as a function of 
$N_{\rm{sig}}$ for the $\sqrt{s}=4.750$~GeV data is shown in 
Figure~\ref{fig:ksksJpsi scan}, where $L$ and $L_0$ are the likelihood values 
for each $N_{\rm{sig}}$ and the maximum likelihood value, respectively.
The upper limit is determined from $\int_{0}^{N_{\rm sig}^{\rm U.L.}}(L/L_0) dx/\int_{0}^{\infty}(L/L_0) dx = 0.9$.

\begin{figure}[htbp]
\begin{center}
\includegraphics[width=0.5\textwidth]{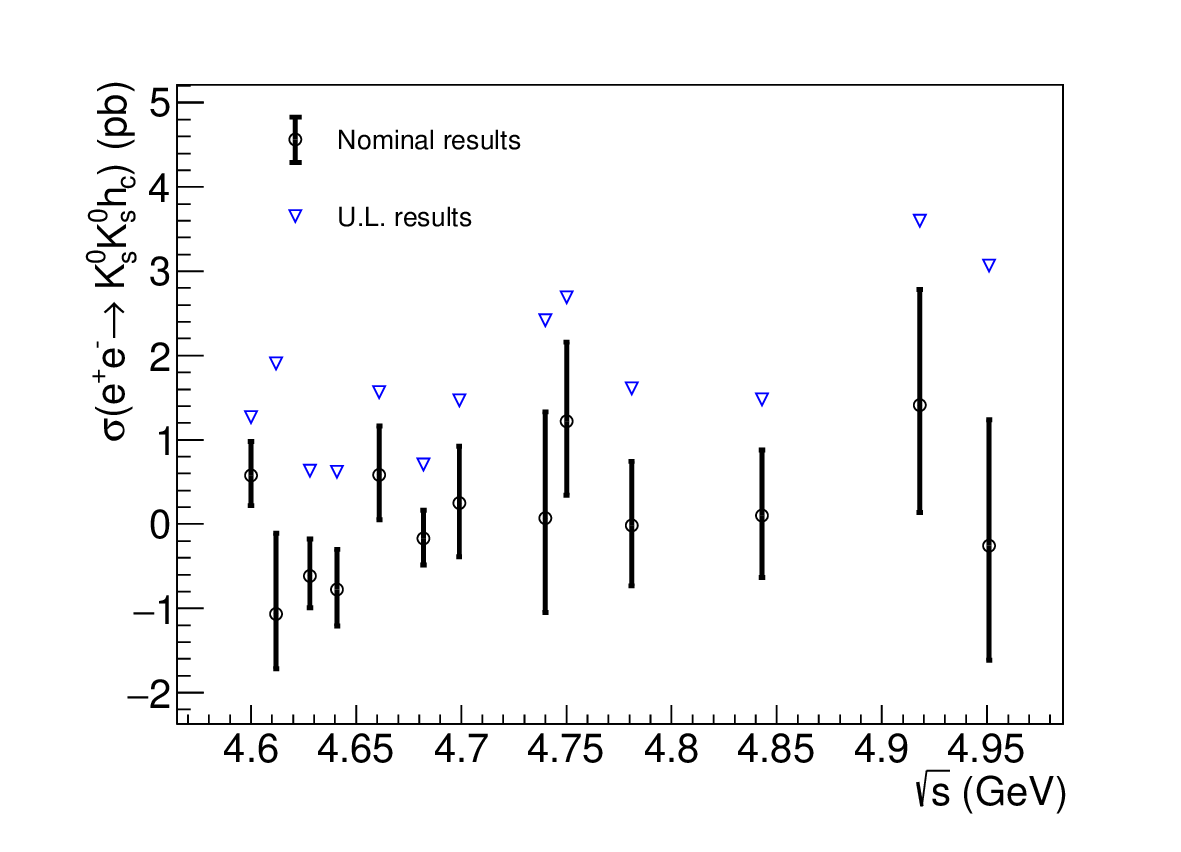}
\end{center}
\caption{
The nominal result (dots with error bars)
and the U.L. (triangular points)
of the Born cross section 
of $\EE\to\ks\ks\hc$.
}
\label{fig:cs}
\end{figure}

The Born cross section $\sigma_{\rm Born}$ is calculated using the formula:
\begin{equation}\label{func:Born cs}
\sigma_{\rm Born}=
\frac{N_{\rm sig}}
{\epsilon {\cal L}(1+\delta)
\delta_{\rm VP}
{\cal B}^2(\ks\to\pp)},
\end{equation}
where $\epsilon$, $\cal{L}$, $(1+\delta)$, $\delta_{\rm VP}$,
and ${\cal B}(\ks\to\pp)$~\cite{PDG}
represent the detection efficiency,
the integrated luminosity, the ISR correction factor,
the vacuum polarization (VP) correction factor~\cite{VP:2010bjp},
and the branching fraction of $\ks\to\pp$, respectively.
The upper limit of $\sigma_{\rm{Born}}$ is similarly determined 
by substituting $N_{\rm{sig}}$ with its upper limit
$N_{\rm{sig}}^{\rm{U.L.}}$.
The numerical results for each data sample are 
listed in Table~\ref{tab:fit result} and shown in 
Figure~\ref{fig:cs}.
In the table, the second error of
$\sigma_{\rm Born}$ represents the systematic uncertainty,
$N_{\rm sig}^{\rm{U.L., nom}}$ is the nominal upper limit of the 
signal yields, and $\sigma_{\rm Born}^{\rm{U.L., sys}}$ reflects the 
most conservative result incorporating systematic uncertainties
that will be discussed in the next section.

\begin{figure*}[htbp]
\begin{center}
\includegraphics[width=0.5\textwidth]{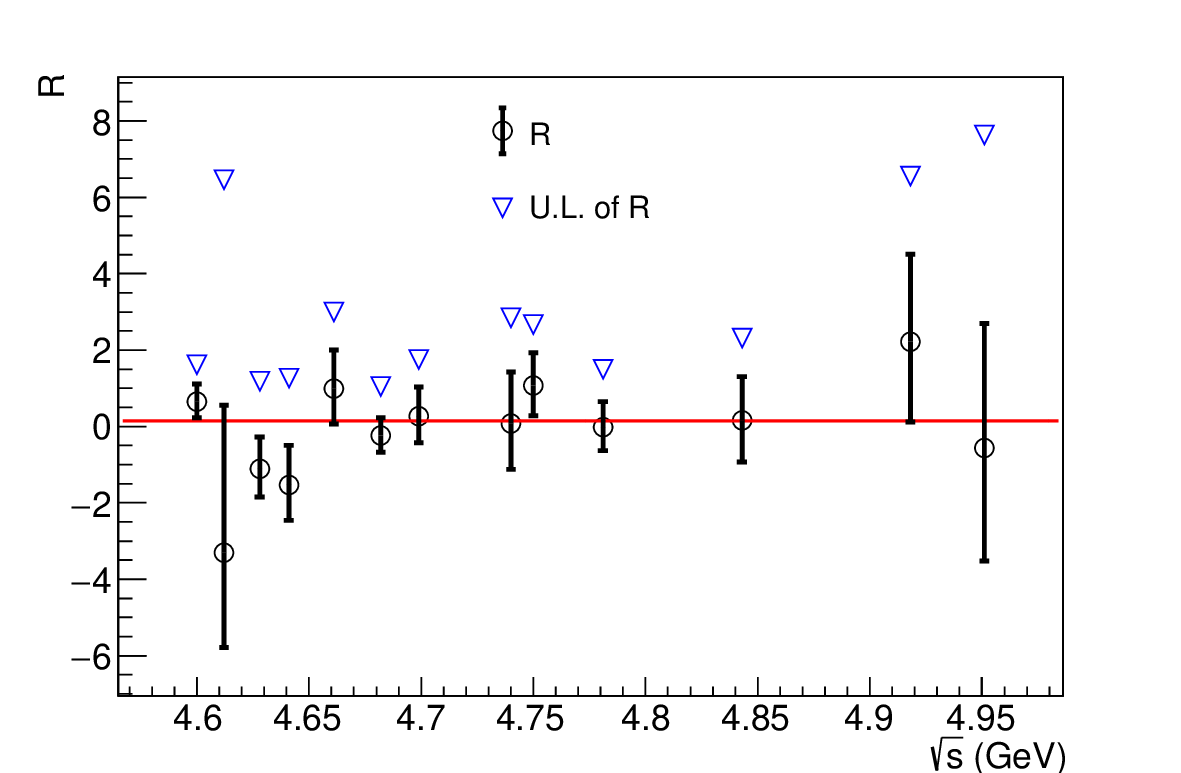}
\end{center}
\caption{
The ratio $R=\frac{\sigma(\EE\to\ks\ks h_c)}{\sigma(\EE\to\ks\ks\jpsi)}$
and its upper limit.
The red line indicates the average value.
}
\label{fig:ratio}
\end{figure*}

The decay of conventional vector charmonium states
into $\hc$ is suppressed due to heavy-quark spin symmetry,
while the size of $\sigma(\EE\to\pp h_c)$ and 
$\sigma(\EE\to\pp\jpsi)$ in the range $4.2<\sqrt{s}<4.4~\gev$
is found similar.
To better understand the nature of $Y$ states above 4.6~GeV,
we calculate the ratio $R=\frac{\sigma(\EE\to\ks\ks h_c)}{\sigma(\EE\to\ks\ks\jpsi)}$.
The values of $\sigma(\EE\to\ks\ks\jpsi)$ are taken from
Ref.~\cite{BESIII:2022joj,BESIII:2023wqy,intro-BESIII-ksksjpsi}.
The results for $\sigma(\EE\to\ks\ks\jpsi)$ and $\sigma(\EE\to K^+K^-\jpsi)$
are combined assuming isospin symmetry.
Since all three measurements are dominated by the statistical uncertainty,
the quadratic sum of the statistical and systematic uncertainties of each 
process is used. The results for $R$ are presented in Figure~\ref{fig:ratio}.
Fitting the ratio with a constant term, the average ratio is determined 
to be $0.15 \pm 0.22$. The upper limit of the cross section ratio 
is calculated by replacing $\sigma(\EE\to\ks\ks h_c)$ with 
$\sigma^{\rm U.L.,sys}(\EE\to\ks\ks h_c)$, determined with the uncertainties 
of the $\sigma(\EE\to\ks\ks\jpsi)$ included as one source of systematic 
uncertainty.

\begin{table*}[htbp]
\caption{
The numerical results of the Born cross section 
$\sigma_{\rm{Born}}$ for each data sample (in unit of pb).
The numbers in brackets are upper limits at the 90\% C.L.
Shown are also the c.m. energies $\sqrt{s}$ (in unit of GeV),
the integrated luminosity $\cal{L}$ (in unit of $\rm{pb}^{-1}$),
the number of signal events $N_{\rm sig}$.
the efficiency $\epsilon$ (in unit of \%),
the VP correction factor $\delta_{\rm VP}$,
the ISR correction factor $(1+\delta)$,
the significance $S$ at each data sample,
the ratio $R=\frac{\sigma(\EE\to\ks\ks h_c)}{\sigma(\EE\to\ks\ks\jpsi)}$.
}
\scriptsize

\label{tab:fit result}
\begin{center}
\begin{tabular}{ccrccccrr}
\hline\hline
$\sqrt{s}$  & $\lum$ & $N_{\rm sig}$ 
($N_{\rm sig}^{\rm{U.L., nom}}$)
& $\epsilon$ & $\delta_{VP}$ & $1+\delta$ & $S$
& $\sigma_{\rm{Born}}$ ($\sigma_{\rm{Born}}^{\rm{U.L., sys}})$~~~~~
& $R$ ($R^{\rm U.L.}$)~~~~~
\\
\hline\hline

4.600 & 587 & $10.3^{+7.0}_{-6.3}(<20.4)$   & 8.4 & 1.055 & 0.720 & 1.7$\sigma$ & 0.58$^{+0.40}_{-0.36}\pm0.08(<1.28)$  & $0.7^{+0.5}_{-0.4}$ $(<1.6)$ \\
4.612 & 104 & $-3.1^{+2.8}_{-1.8}(<4.8)$    & 7.6 & 1.055 & 0.731 & $-$         &-1.07$^{+0.96}_{-0.65}\pm0.16(<1.91)$  &$-3.3^{+3.9}_{-2.5}$ $(<6.5)$ \\
4.628 & 522 & $-9.4^{+6.6}_{-5.8}(<8.3)$    & 7.9 & 1.054 & 0.741 & $-$         &-0.61$^{+0.44}_{-0.38}\pm0.08(<0.63)$  &$-1.1^{+0.9}_{-0.7}$ $(<1.2)$ \\
4.641 & 552 &$-12.8^{+7.8}_{-7.0}(<8.8)$    & 8.0 & 1.054 & 0.745 & $-$         &-0.78$^{+0.48}_{-0.43}\pm0.09(<0.62)$  &$-1.5^{+1.0}_{-0.9}$ $(<1.3)$ \\
4.661 & 529 &  $9.8^{+9.7}_{-8.9}(<23.8)$   & 8.3 & 1.054 & 0.756 & 1.1$\sigma$ & 0.59$^{+0.58}_{-0.54}\pm0.07(<1.57)$  & $1.0^{+1.0}_{-0.9}$ $(<3.0)$ \\
4.682 &1667 & $-9.3^{+18.2}_{-17.3}(<25.9)$ & 8.6 & 1.054 & 0.764 & $-$         &-0.17$^{+0.33}_{-0.32}\pm0.02(<0.71)$  &$-0.2^{+0.5}_{-0.4}$ $(<1.1)$ \\
4.699 & 536 &  $4.3^{+11.8}_{-11.0}(<23.0)$ & 8.4 & 1.055 & 0.769 & 0.4$\sigma$ & 0.25$^{+0.68}_{-0.63}\pm0.03(<1.48)$  & $0.3^{+0.8}_{-0.7}$ $(<1.8)$ \\
4.740 & 165 &  $0.4^{+7.4}_{-6.6}(<13.4)$   & 9.1 & 1.055 & 0.781 & 0.1$\sigma$ & 0.07$^{+1.26}_{-1.12}\pm0.03(<2.42)$  & $0.1^{+1.4}_{-1.2}$ $(<2.8)$ \\
4.750 & 367 & $16.4^{+12.5}_{-11.7}(<33.8)$ & 9.2 & 1.055 & 0.782 & 1.4$\sigma$ & 1.22$^{+0.94}_{-0.88}\pm0.12(<2.71)$  & $1.1^{+0.9}_{-0.8}$ $(<2.7)$ \\
4.781 & 511 & $-0.3^{+14.6}_{-13.7}(<24.7)$ & 9.4 & 1.055 & 0.788 & $-$         &-0.02$^{+0.76}_{-0.72}\pm0.01(<1.62)$  & $0^{+0.7}_{-0.6}$ $(<1.5)$ \\
4.843 & 525 &  $2.1^{+15.5}_{-14.7}(<27.7)$ & 9.3 & 1.056 & 0.802 & 0.1$\sigma$ & 0.11$^{+0.78}_{-0.74}\pm0.02(<1.48)$  & $0.2^{+1.1}_{-1.1}$ $(<2.3)$ \\
4.918 & 208 & $11.7^{+11.2}_{-10.5}(<27.9)$ & 9.7 & 1.056 & 0.814 & 1.1$\sigma$ & 1.41$^{+1.37}_{-1.28}\pm0.19(<3.62)$  & $2.2^{+2.3}_{-2.1}$ $(<6.6)$ \\
4.951 & 159 & $-1.6^{+9.5}_{-8.6}(<15.6)$   & 9.6 & 1.056 & 0.818 & $-$         &-0.26$^{+1.49}_{-1.36}\pm0.03(<3.10)$  &$-0.6^{+3.3}_{-3.0}$ $(<7.7)$ \\

\hline
\end{tabular}
\end{center}
\end{table*}

\section{Systematic uncertainty}

The uncertainties in the measured Born cross sections
arise from various sources,
which contribute either as multiplicative or additive terms.
The multiplicative terms include
integrated luminosity,
input branching fractions,
and detection efficiency.
Additive terms stem from
the determination of $N_{\rm sig}$
and the line shape of the cross section.
The uncertainties are combined in quadrature to calculate the total systematic uncertainty,
assuming they are independent.
For the upper limit, 
the uncertainty from multiplicative terms,
summarized in Table~\ref{table:sys multiplicative},
is incorporated by convoluting a Gaussian function
into the likelihood distribution.
The likelihood distribution is derived from the most
conservative result based on the additive terms.

\begin{table}[htbp]
\caption{Summary of multiplicative systematic uncertainties.}
\label{table:sys multiplicative}
\begin{center}
\begin{tabular}{cc}
\hline\hline
Source & Uncertainty (\%)  \\\hline
Luminosity         & 1.0    \\
$\mathcal{B}(\hc\to\gamma\etac)$   &  4.3  \\
Photon reconstruction    & 1.0  \\
$\ks$ reconstruction & 2.6 $-$ 8.8 \\
$\etac$ mass window & 1.0 \\
Simulation model & 6.0  \\
\hline
Total  & 8.1 $-$ 11.7 \\
\hline
\hline
\end{tabular}
\end{center}
\end{table}

The integrated luminosity is measured by selecting Bhabha scattering events,
with an associated uncertainty of $1.0\%$
\cite{BESIII-lumi-yifan, cms-lumi-round1314}.
Since the processes $\hc\to\gamma\etac$
and $\hc\to\rm{non-}(\gamma\etac)$ are both considered as signal,
the uncertainty from ${\cal B} (\hc\to\gamma\etac)$,
for which the best measurement yields
${\cal B} (\hc\to\gamma\etac)=(57.66^{+3.62}_{-3.50}\pm0.58)\%$~\cite{BESIII:2022tfo},
is estimated by modifying the value by $\pm1\sigma$.
The resulting change in the detection efficiency $\epsilon$ of 4.3\%
is taken as the systematic uncertainty.

The uncertainty from photon reconstruction is studied using
control samples of $\jpsi\to\rho\pi^0$
and $\EE\to\gamma\gamma$, and is determined to be
1\% per photon~\cite{phton detection efficiency}.
The uncertainty from $\ks$ reconstruction
is assessed by selecting the decay $\jpsi\to
K^{*}{}^{\pm}\bar{K}^{\mp}$ as a control sample.
The reconstruction efficiency 
difference between data and MC sample as a function of the 
momentum of $\ks$ ($p_{\ks}$) is provided.
By weighing the efficiency difference 
according to $p_{\ks}$ in the signal process
across various data samples,
the systematic uncertainty from $\ks$ is 
estimated to range from $4.4\%$ to $1.3\%$ per $\ks$
for $\sqrt{s}$ ranging from $4.600~\gev$ to $4.951~\gev$.
Uncertainties from the parameters of $\etac$ in the MC sample
are estimated by varying the mass and width separately
by $\pm 1\sigma$, resulting in an uncertainty of 0.5\%,
where $\sigma$ values are cited from PDG~\cite{PDG}.
The uncertainty from the line shape of $\etac$~\cite{Anashin:2010dh}
used in MC samples is estimated by incorporating the missing term ($E_{\gamma}^3$),
resulting in a difference in detection efficiencies of 0.8\%.
Consequently, the total uncertainty from the $\etac$ mass window is 1\%.

The systematic uncertainty caused by the simulation model
is estimated by producing MC samples of the processes 
$\EE\to \ks Z_{cs}(4220)$,
$\EE\to f_0(1370)h_c$, 
or $\EE\to f_2(1270)h_c$.
The mass and width of $Z_{cs}(4220)$ are fixed 
according to the result reported by LHCb~\cite{LHCb:2021uow}.
We further simulate the sample with $M(Z_{cs}(4220))$
shifted by $+50~\mevcc$ since the 
mass of $Z_{cs}^0$ is expected to be larger than 
$Z_{cs}^\pm$~\cite{Wan:2020oxt}.
The efficiency difference compared to the nominal value
is 6\% and is taken as the systematic uncertainty.
The systematic uncertainty caused by the limited statistics
of MC sample is calculated with 
$\delta \epsilon = \sqrt{\frac{\epsilon (1-\epsilon)}{N_{\rm gen}}}$,
where $N_{\rm gen}$ represents the number of generated events.
The combined uncertainty for $\hc\to (\gamma\etac)$
and $\hc\to\rm{non-}(\gamma\etac)$
processes is calculated error propagation formula.
The uncertainty is calculated to be 1\%.

In the fit to the $M_{\ksks}^{\rm rec}$ distribution,
the uncertainty due to the resolution difference
between data and MC sample
is estimated by generating ensemble of pseudoexperiments with
parameters modified by 0 or $\pm1\sigma$.
This results in one set of nominal shape MC samples
and eight sets for the modified shapes.
The ensemble of pseudoexperiments
are then fitted using the nominal line shape.
The systematic uncertainty is determined by comparing the fit
results obtained from toy MC samples based on the nominal and modified shapes,
which is found to range within $\delta(N_{\rm sig})=(0.0,0.5)$
for different data samples.
For the upper limit, the uncertainty is estimated 
by repeating the scan with the modified shape.
The uncertainty from the fit range is assessed
following the discussion in~\cite{Barlow1,Barlow2}.
The lower and upper boundaries of the fit range are modified
separately by $\pm 5~\mevcc$ and $\pm 10~\mevcc$.
The study indicates that the effect of the fit range is negligible.
The systematic uncertainty arising from the background model
is assessed by repeating the fit with 
the second-order Chebyshev function modified to a third-order Chebyshev function.
The change in the value of $(-\ln L)$,
which measures the improvement,
is found to be negligible.
The likelihood scan is repeated with the modified background shape,
and the largest U.L. is taken as a conservative estimation.

The input cross section line shape affects not only the
ISR correction factor $(1+\delta)$ and the detection efficiency
but also the signal shape.
In the nominal result, the cross section is assumed to follow
the three-body decay phase space factor.
We then modify it to reflect the measured cross section line shape 
of the $\EE\to\ksks\jpsi$ process~\cite{intro-BESIII-ksksjpsi}
and take the difference in the fit to estimate the effect.
This uncertainty on the cross section is determined
to range within $\delta(\sigma)=(0.01,0.15)~{\rm pb}$
for the data samples.

\section{Summary}

In summary, we search for the process $\EE\to\ksks\hc$ 
using 13 data samples collected by the BESIII detector
at $\sqrt{s}$ ranging from $4.600~\gev$ to $4.951~\gev$.
The significance of the signal process is found 
to be below $2\sigma$ for each c.m. energy.
The upper limits of the cross section for each data sample are determined 
at the 90\% C.L. based on the current statistics.
There appears to be a slight enhancement of the cross section around 4.75 GeV,
but no definitive conclusions can be made regarding whether
$Y(4710)$ or $Y(4750)$ decay into $\ksks\hc$.
The ratio $R=\frac{\sigma(\EE\to\ks\ks h_c)}{\sigma(\EE\to\ks\ks\jpsi)}$,
calculated by combining the measurements of
$\sigma(\EE\to\ks\ks\jpsi)$ and
$\sigma(\EE\to K^+K^-\jpsi)$,
yields an average value of $0.15 \pm 0.22$.
This ratio suggests that the decay of the $Y$ states
into $h_c$ is substantially smaller than into $\jpsi$,
which differs from the ratio
$\frac{\sigma(\EE\to\pp h_c)}{\sigma(\EE\to\pp\jpsi)}$
in the range $4.2<\sqrt{s}<4.4~\gev$.
The study of $\EE\to\kk\hc$ will be 
presented in a separate paper.
The planned upgrade of the BEPCII~\cite{BESIII:2020nme}
and the anticipated increase
in statistical data in the near future
will enable more precise
results for understanding the $Y(4710)$ and $Y(4750)$ states.

\acknowledgments
The BESIII Collaboration thanks the staff of BEPCII and the IHEP computing center for their strong support. This work is supported in part by National Key R\&D Program of China under Contracts Nos. 2020YFA0406300, 2020YFA0406400, 2023YFA1606000; National Natural Science Foundation of China (NSFC) under Contracts Nos. 11635010, 11735014, 11935015, 11935016, 11935018, 12025502, 12035009, 12035013, 12061131003, 12192260, 12192261, 12192262, 12192263, 12192264, 12192265, 12221005, 12225509, 12235017, 12361141819, 12375070; the Chinese Academy of Sciences (CAS) Large-Scale Scientific Facility Program; the CAS Center for Excellence in Particle Physics (CCEPP); Joint Large-Scale Scientific Facility Funds of the NSFC and CAS under Contract No. U2032108; Shanghai Leading Talent Program of Eastern Talent Plan under Contract No. JLH5913002; 100 Talents Program of CAS; The Institute of Nuclear and Particle Physics (INPAC) and Shanghai Key Laboratory for Particle Physics and Cosmology; German Research Foundation DFG under Contracts Nos. FOR5327, GRK 2149; Istituto Nazionale di Fisica Nucleare, Italy; Knut and Alice Wallenberg Foundation under Contracts Nos. 2021.0174, 2021.0299; Ministry of Development of Turkey under Contract No. DPT2006K-120470; National Research Foundation of Korea under Contract No. NRF-2022R1A2C1092335; National Science and Technology fund of Mongolia; National Science Research and Innovation Fund (NSRF) via the Program Management Unit for Human Resources \& Institutional Development, Research and Innovation of Thailand under Contracts Nos. B16F640076, B50G670107; Polish National Science Centre under Contract No. 2019/35/O/ST2/02907; Swedish Research Council under Contract No. 2019.04595; The Swedish Foundation for International Cooperation in Research and Higher Education under Contract No. CH2018-7756; U. S. Department of Energy under Contract No. DE-FG02-05ER41374

\section{Appendix}

The following plots display the fit results for each data sample.
Each figure consists of three panels:
the left panel shows the fit to the MC sample,
the middle panel shows the fit to the data,
and the right panel displays the likelihood scan.
In the left panel, the black dots with error bars
and the blue histogram represent the MC sample 
and the total fit,, respectively.
In the middle panel, the black dots with error bars,
the blue histogram, the blue dashed histogram,
and the red histogram represent the data sample,
the total fit, the background contribution,
and the signal.
The $\chi$ distribution is presented in the bottom panels for the
fit to the MC and data samples.
The red vertical line in the right panel represents the upper limit at 90\% C.L.

\begin{figure*}[htbp]
\begin{center}
\includegraphics[width=0.63\textwidth]{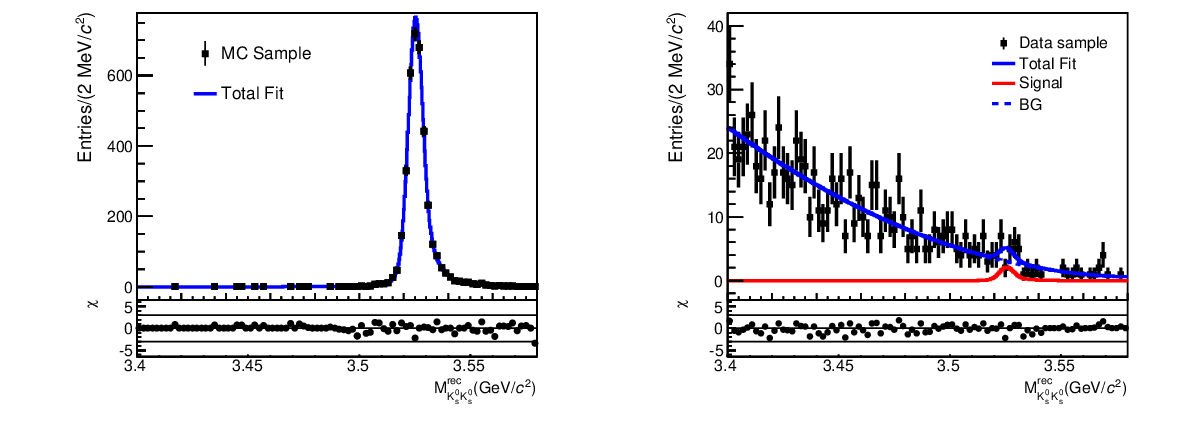}
\includegraphics[width=0.33\textwidth]{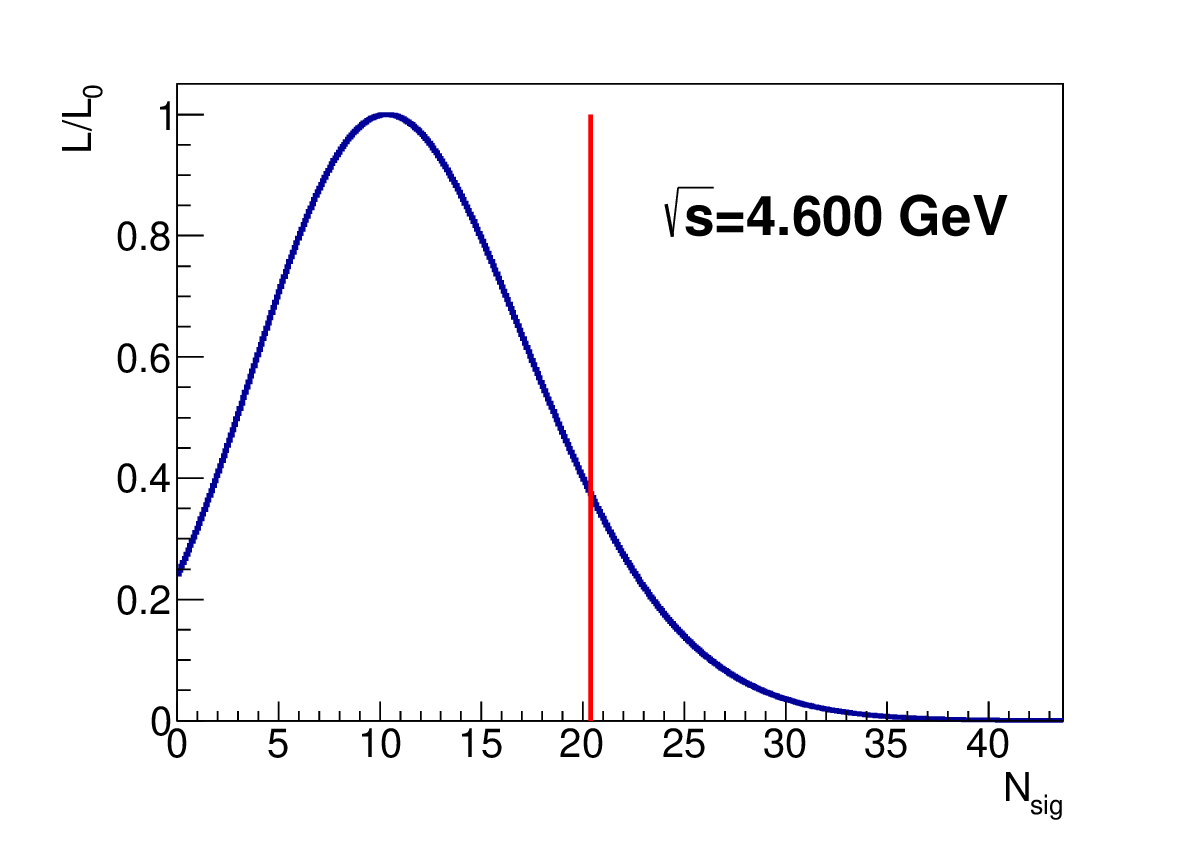}
\\
\includegraphics[width=0.63\textwidth]{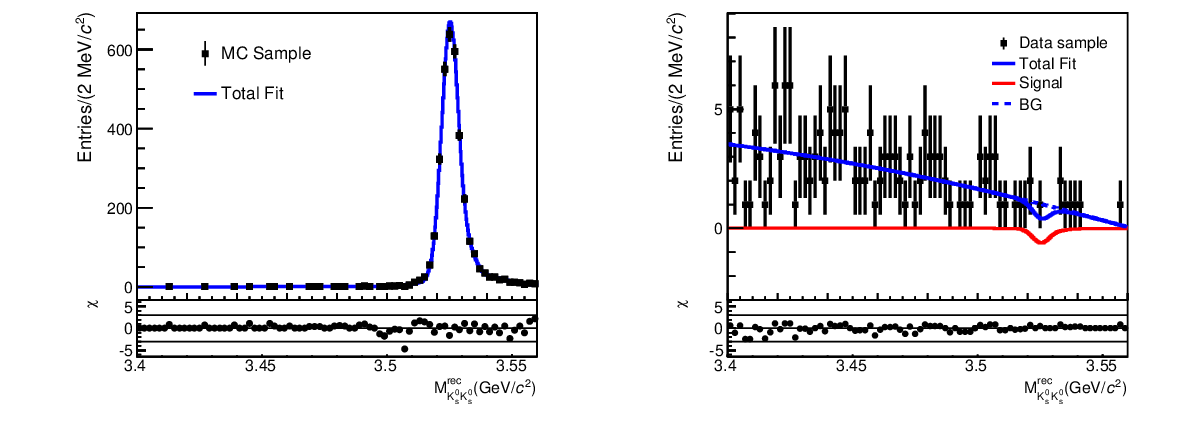}
\includegraphics[width=0.33\textwidth]{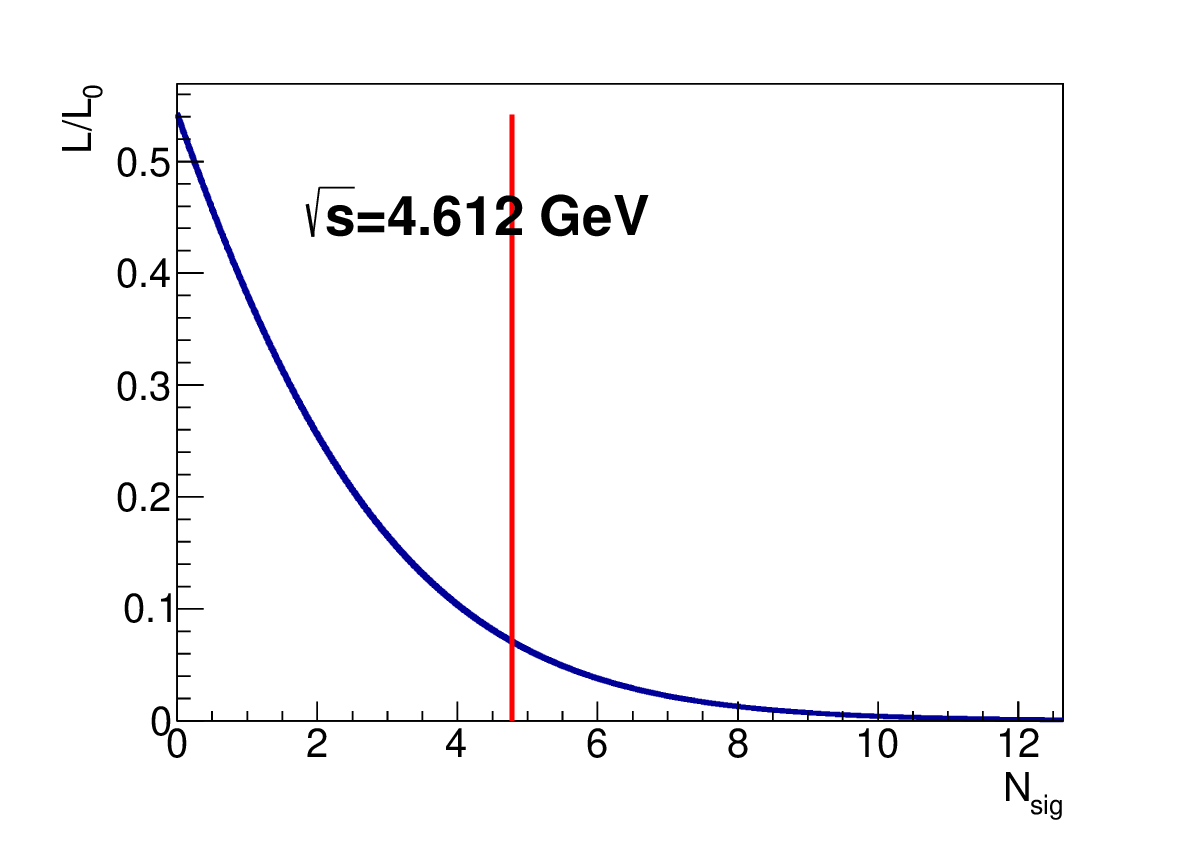}
\\
\includegraphics[width=0.63\textwidth]{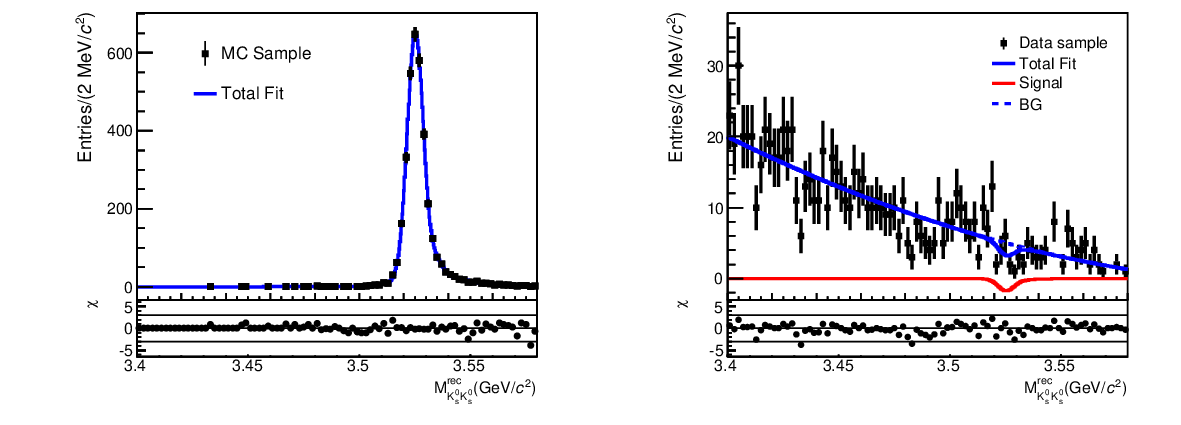}
\includegraphics[width=0.33\textwidth]{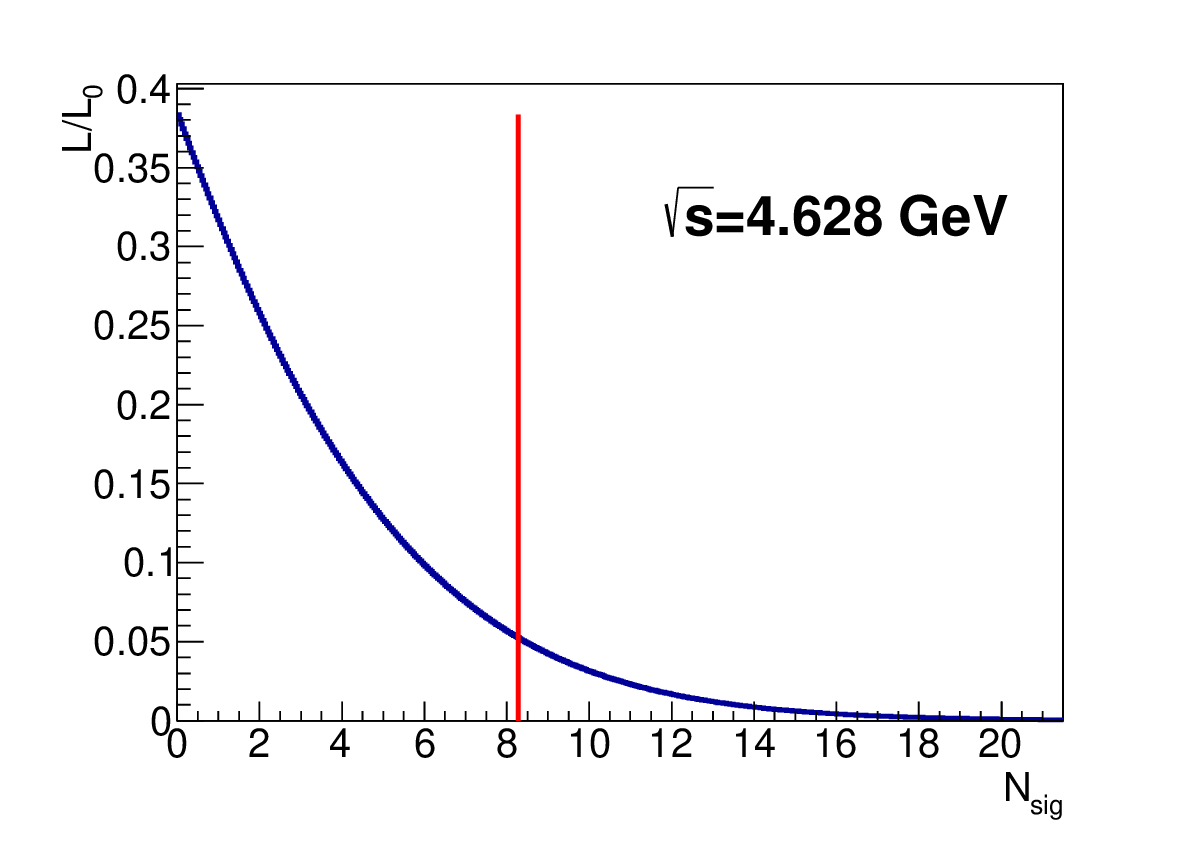}
\\
\includegraphics[width=0.63\textwidth]{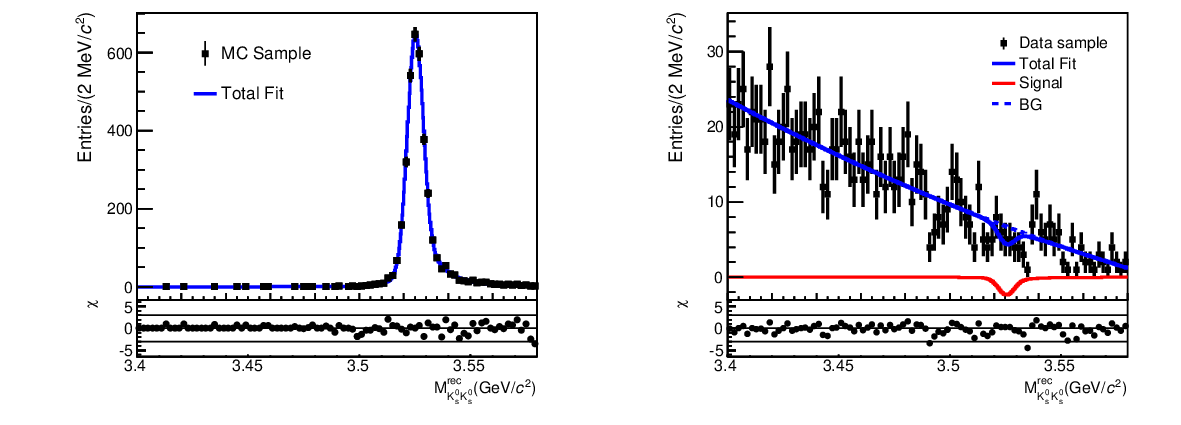}
\includegraphics[width=0.33\textwidth]{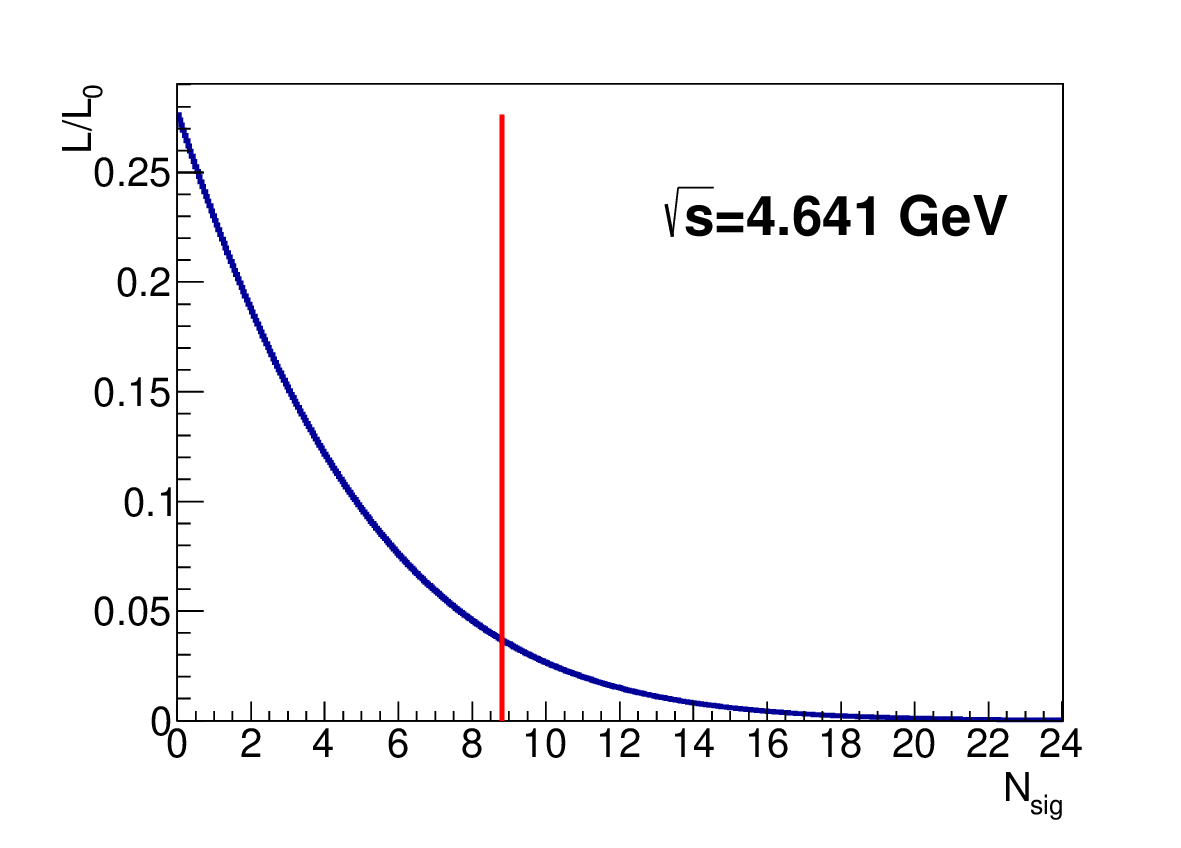}
\\
\includegraphics[width=0.63\textwidth]{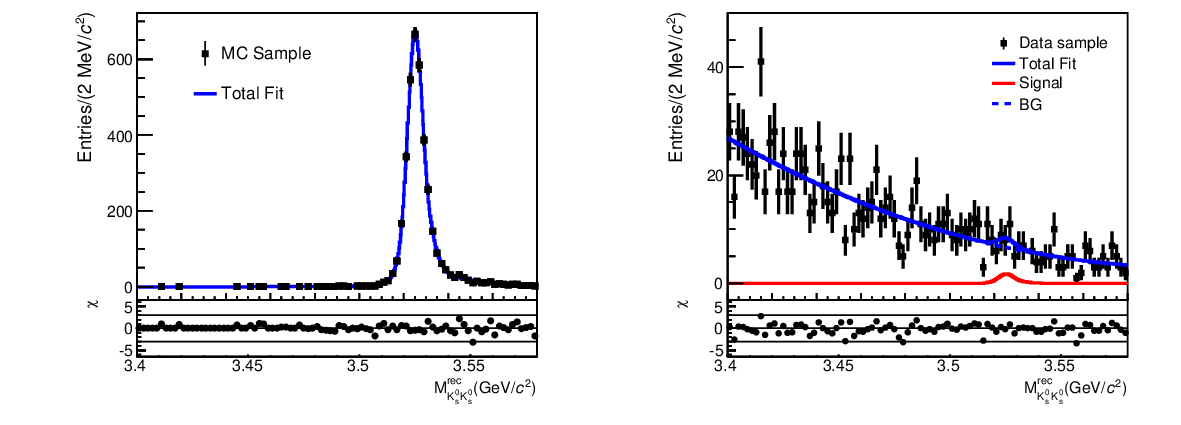}
\includegraphics[width=0.33\textwidth]{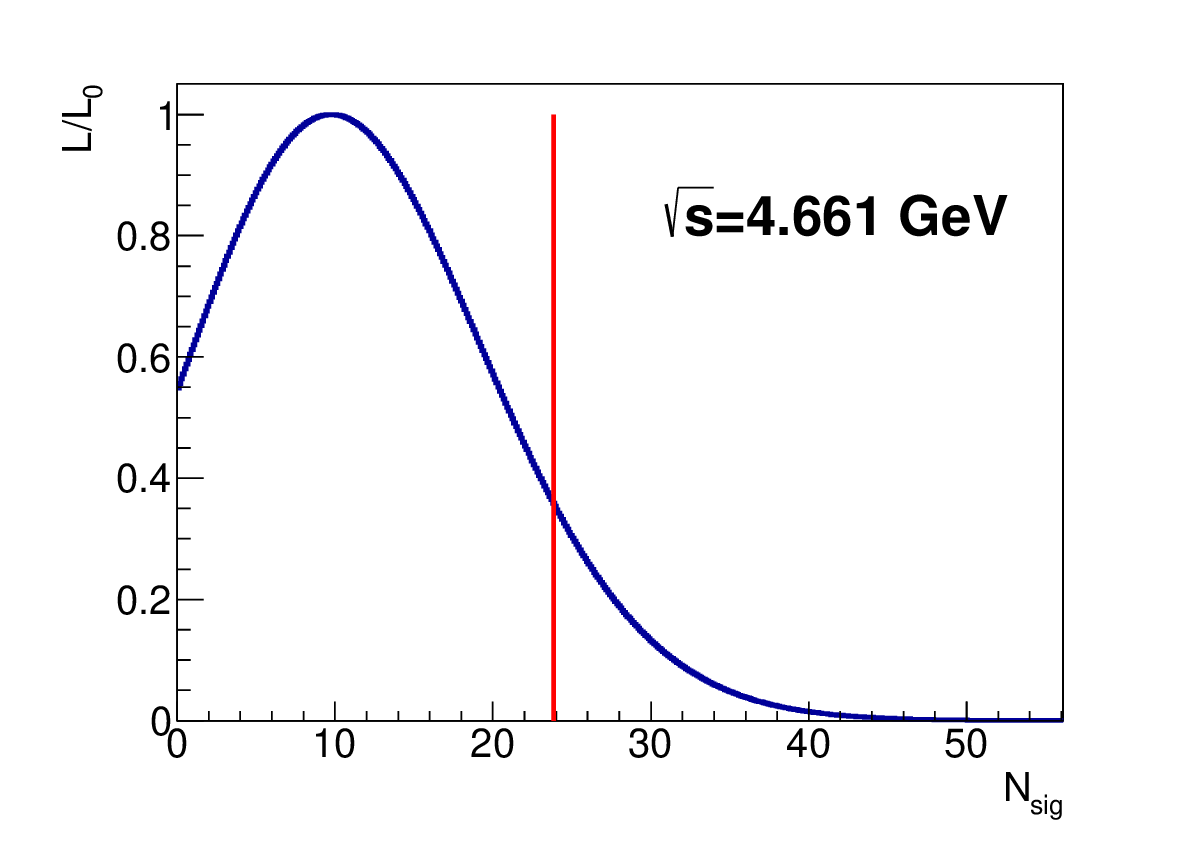}
\end{center}
\caption{Fit and scan results at $\sqrt{s}=4.600-4.661$~GeV.}
\label{fig:kskshc fit 4600}
\end{figure*}

\begin{figure*}[htbp]
\begin{center}
\includegraphics[width=0.63\textwidth]{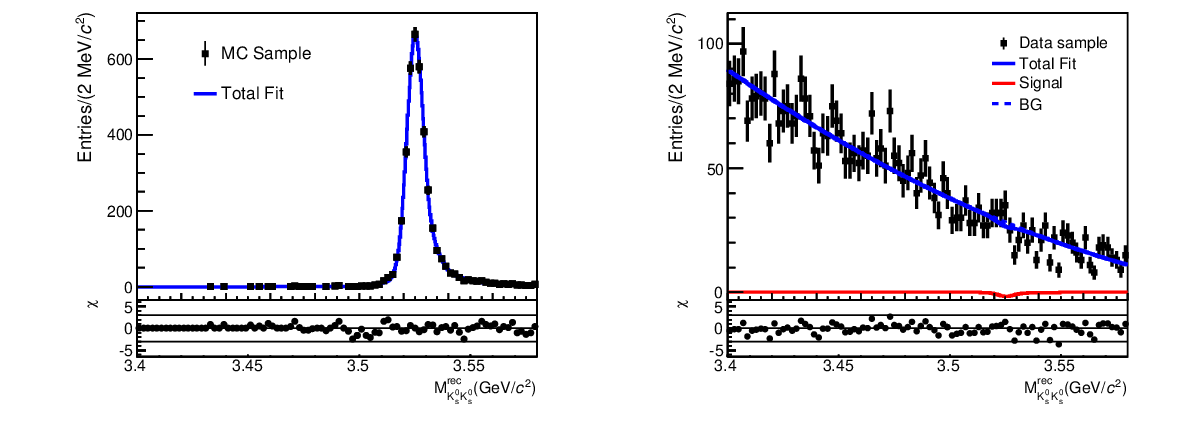}
\includegraphics[width=0.33\textwidth]{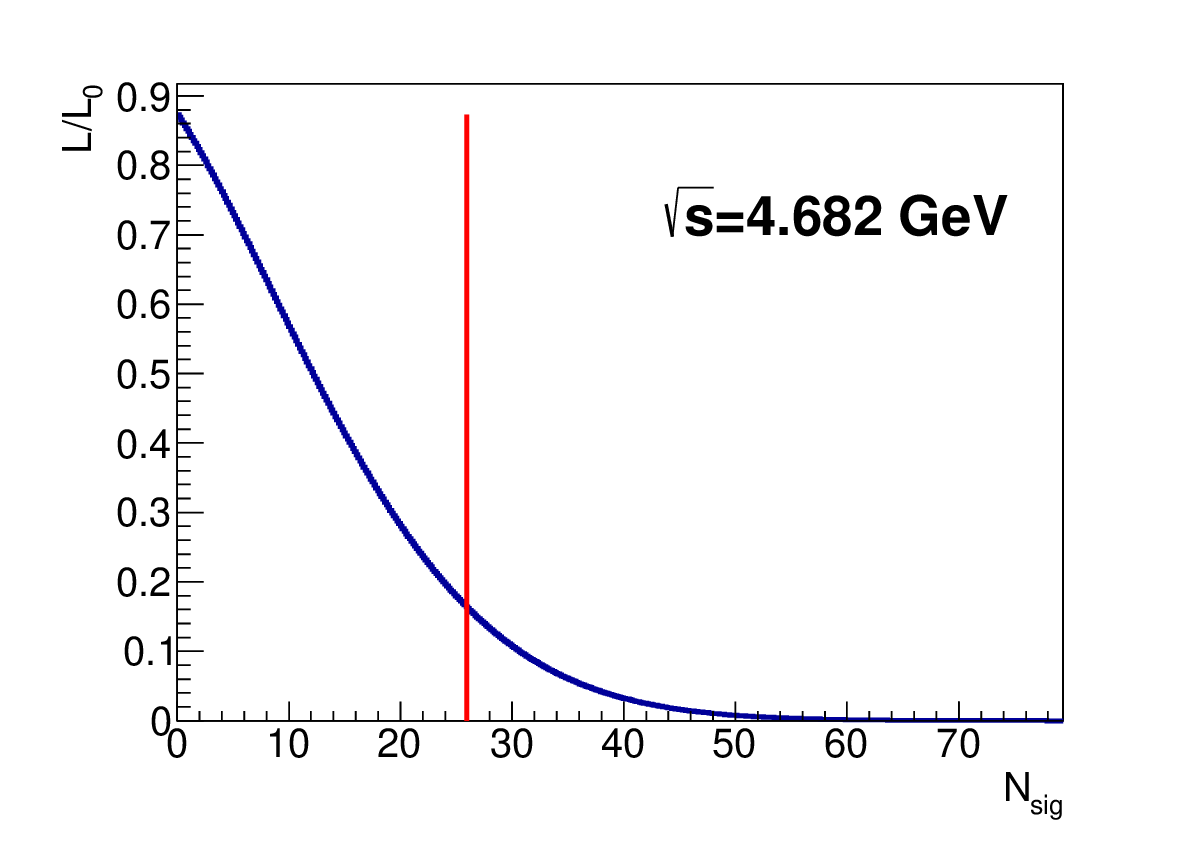}
\\
\includegraphics[width=0.63\textwidth]{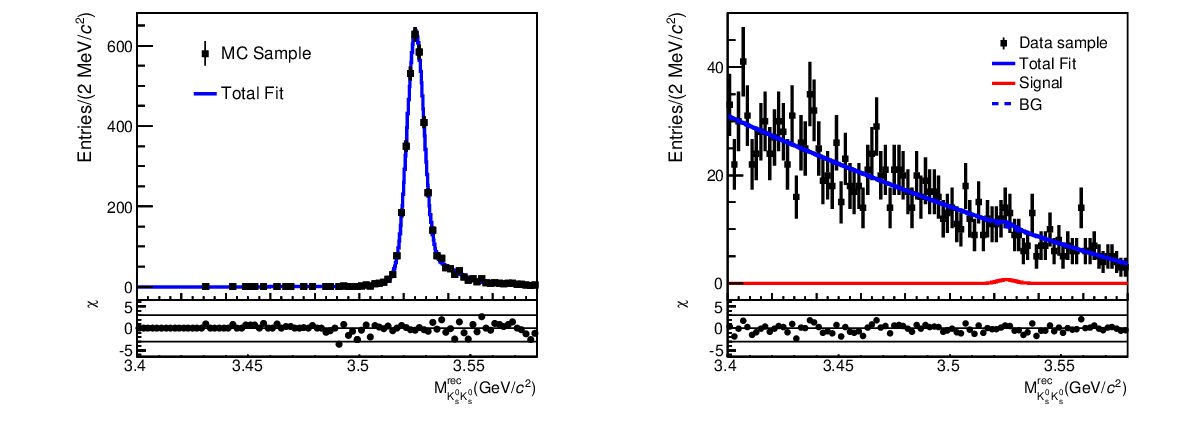}
\includegraphics[width=0.33\textwidth]{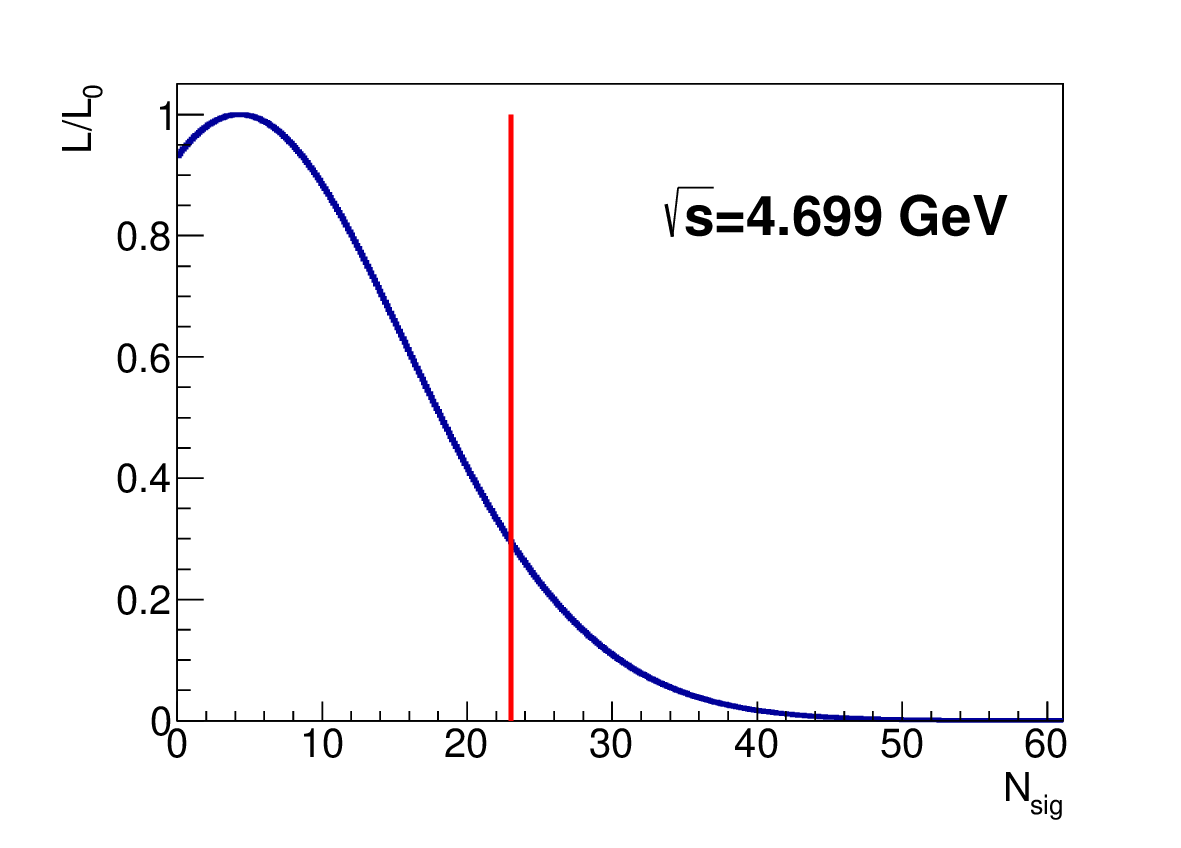}
\\
\includegraphics[width=0.63\textwidth]{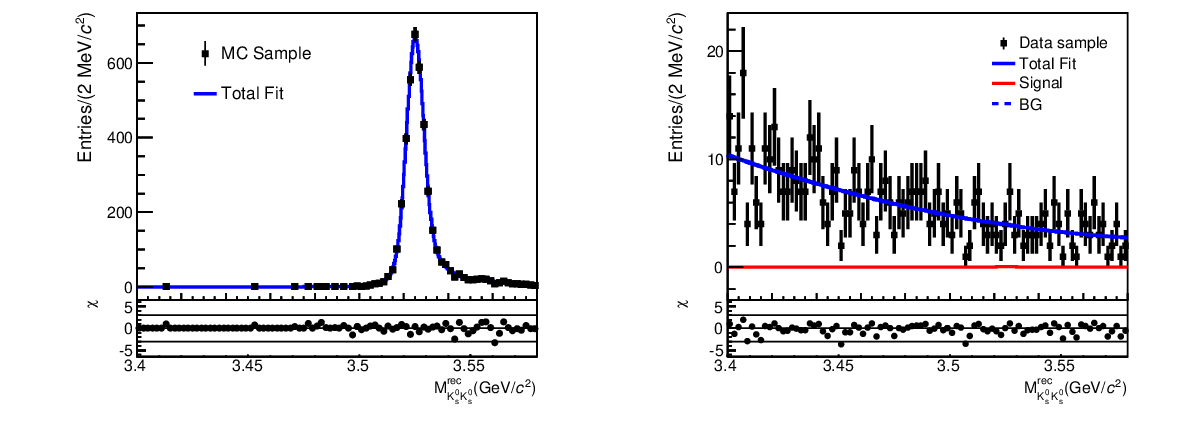}
\includegraphics[width=0.33\textwidth]{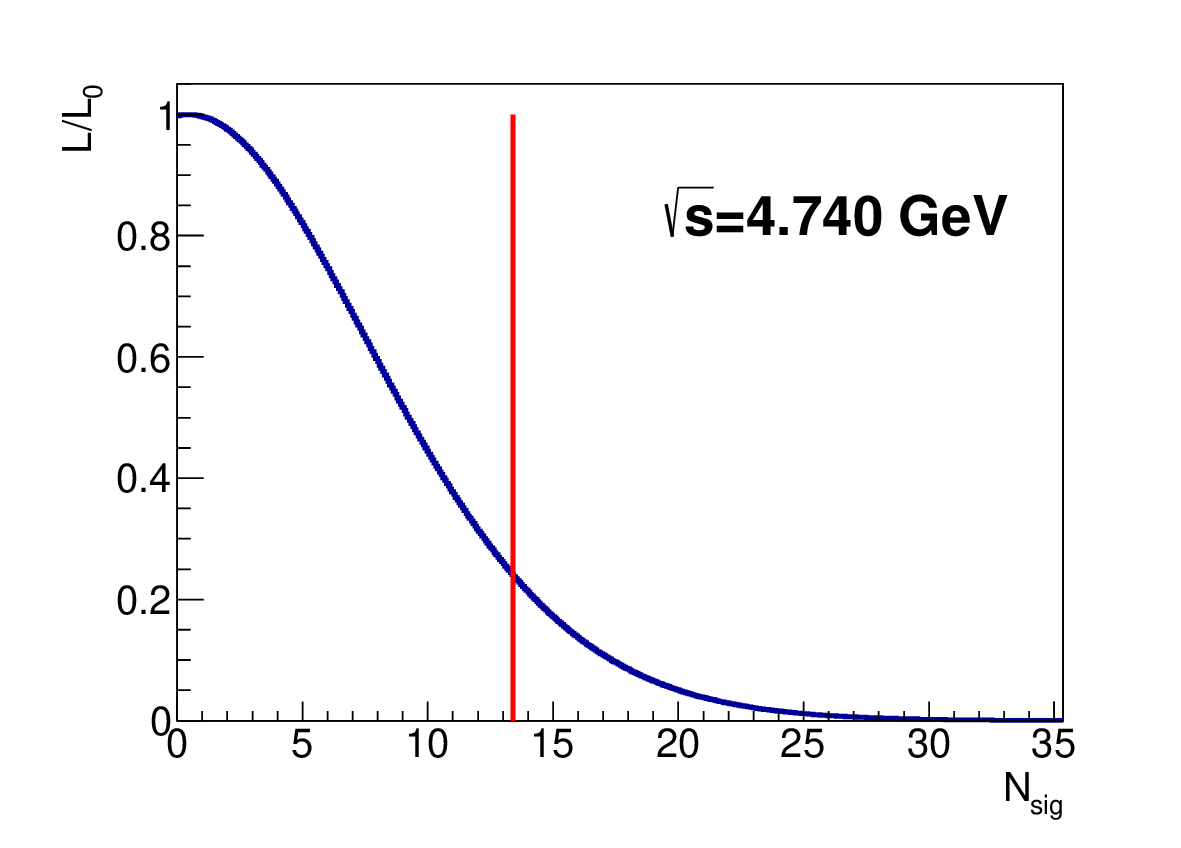}
\\
\includegraphics[width=0.63\textwidth]{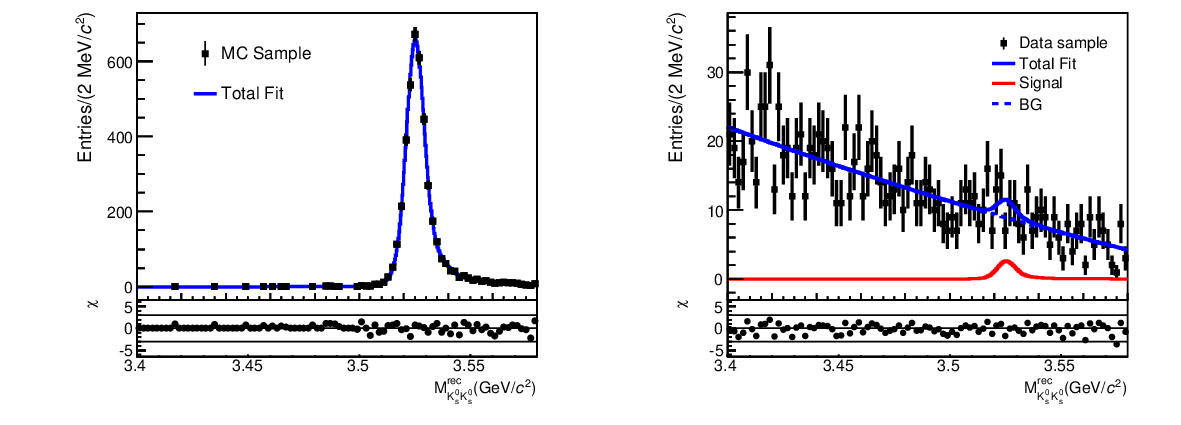}
\includegraphics[width=0.33\textwidth]{pic_supp/fit_scan_4750}
\\
\includegraphics[width=0.63\textwidth]{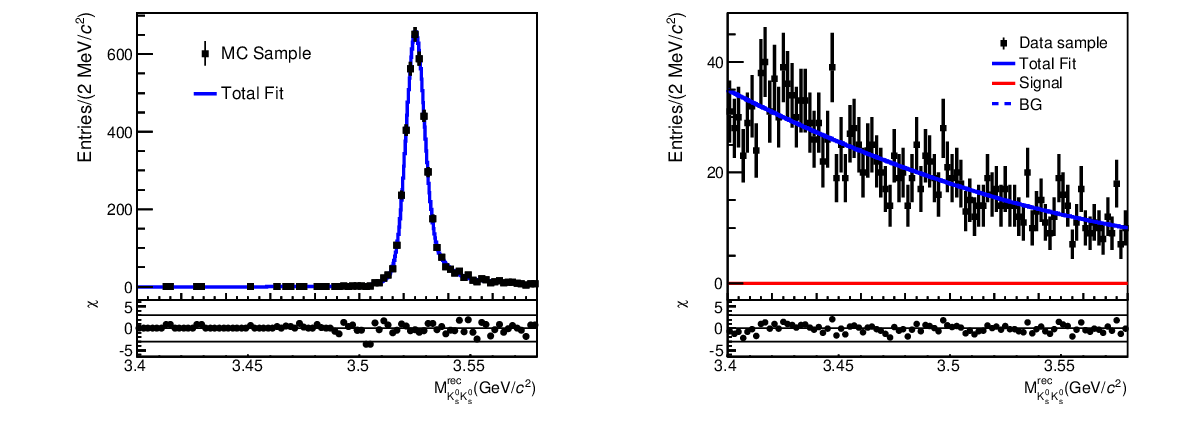}
\includegraphics[width=0.33\textwidth]{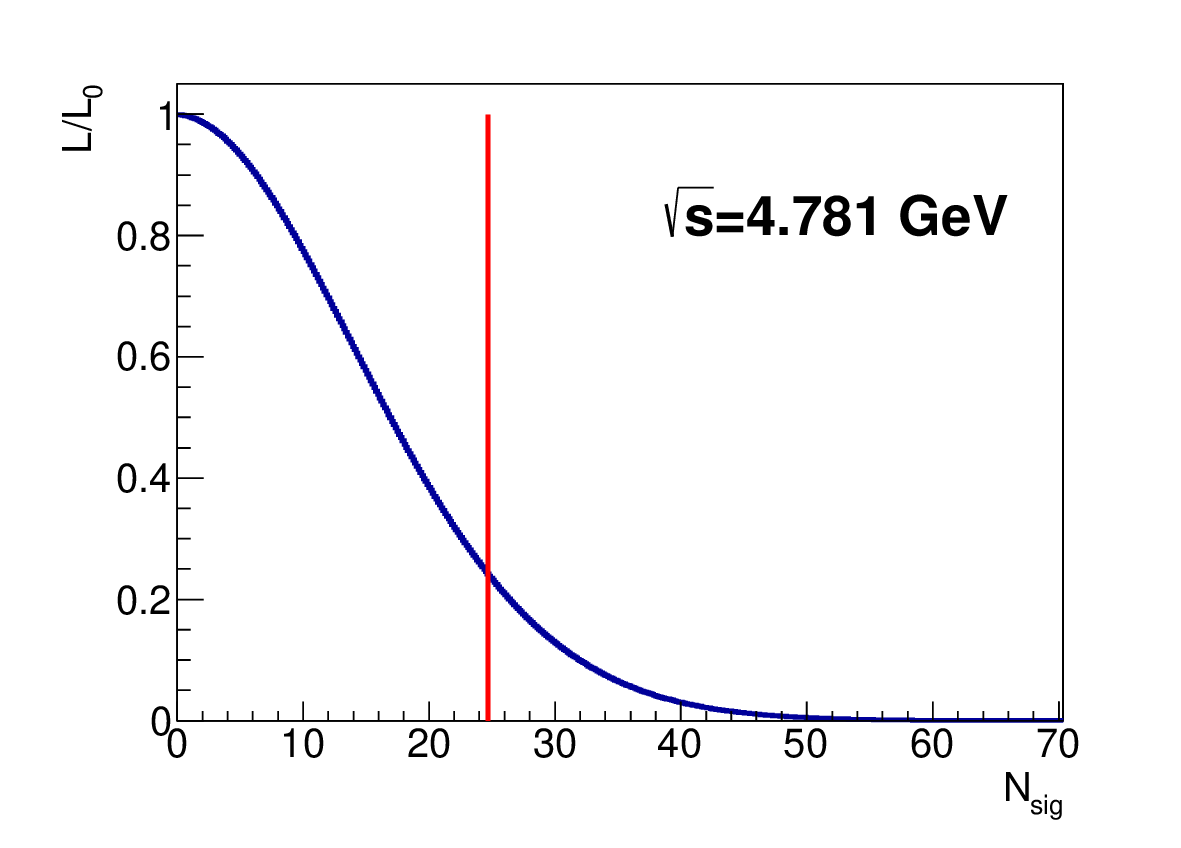}
\end{center}
\caption{Fit and scan results at $\sqrt{s}=4.682-4.781$~GeV.}
\label{fig:kskshc fit 4680}
\end{figure*}

\begin{figure*}[htbp]
\begin{center}
\includegraphics[width=0.63\textwidth]{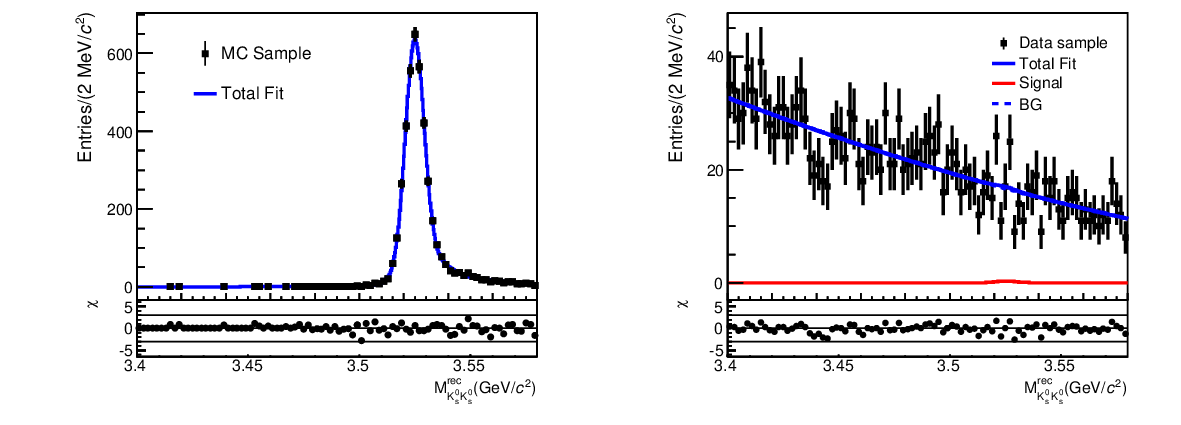}
\includegraphics[width=0.33\textwidth]{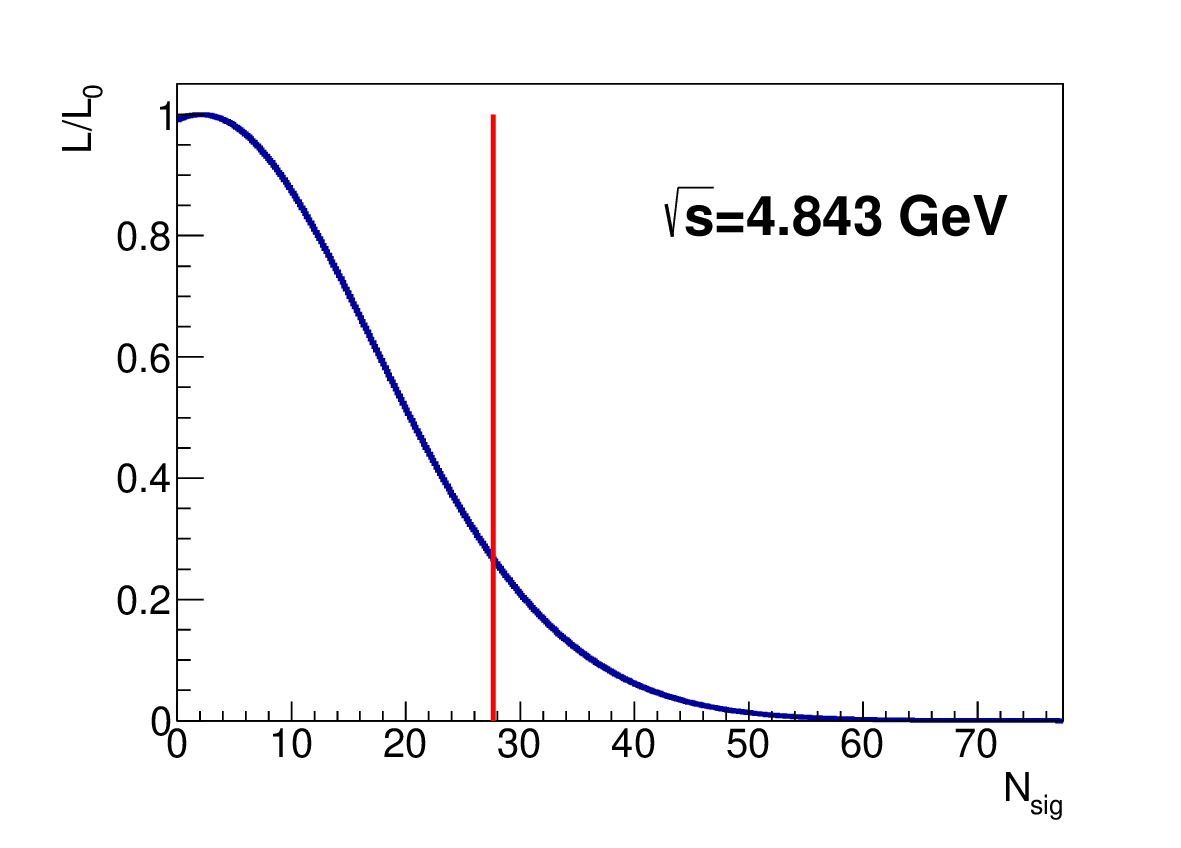}
\\
\includegraphics[width=0.63\textwidth]{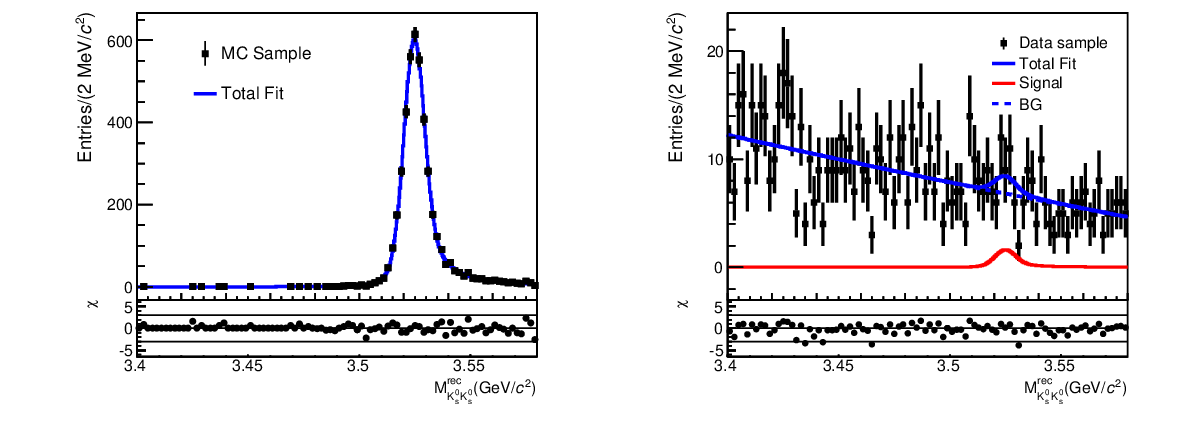}
\includegraphics[width=0.33\textwidth]{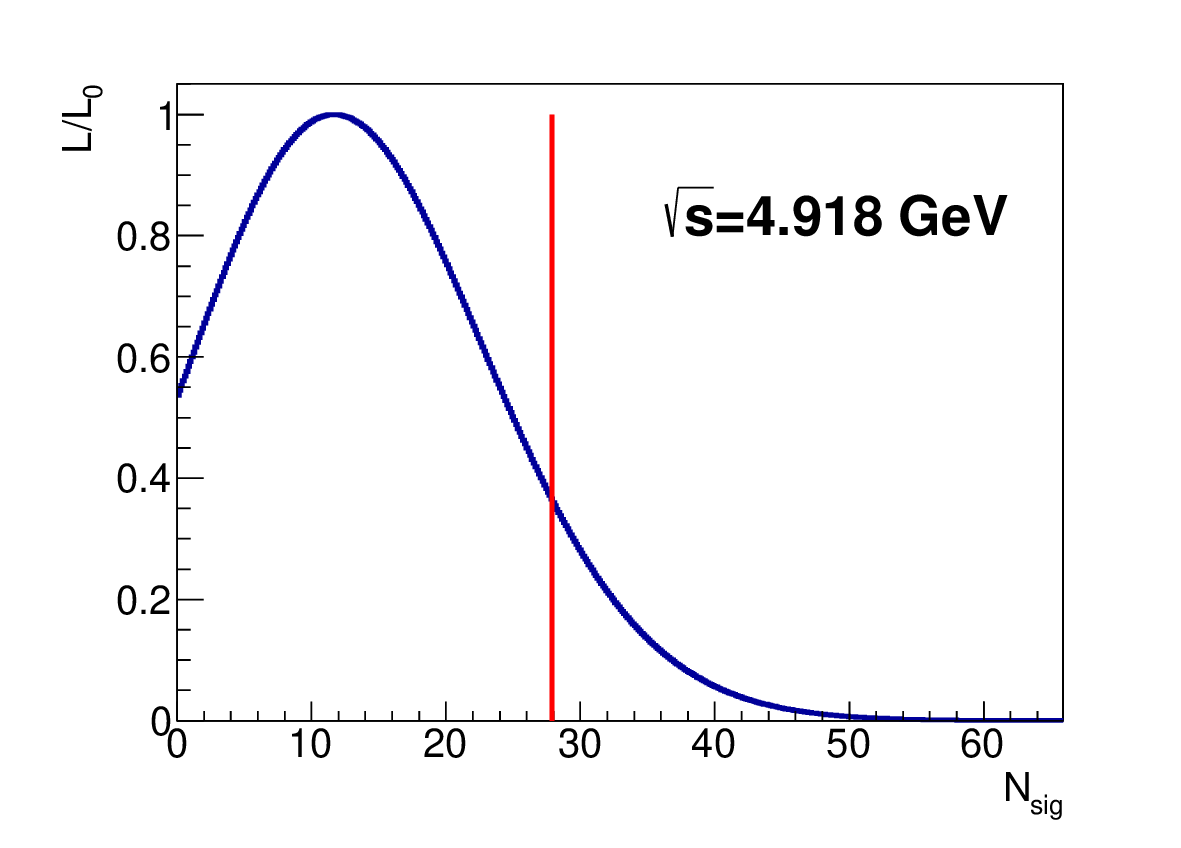}
\\
\includegraphics[width=0.6\textwidth]{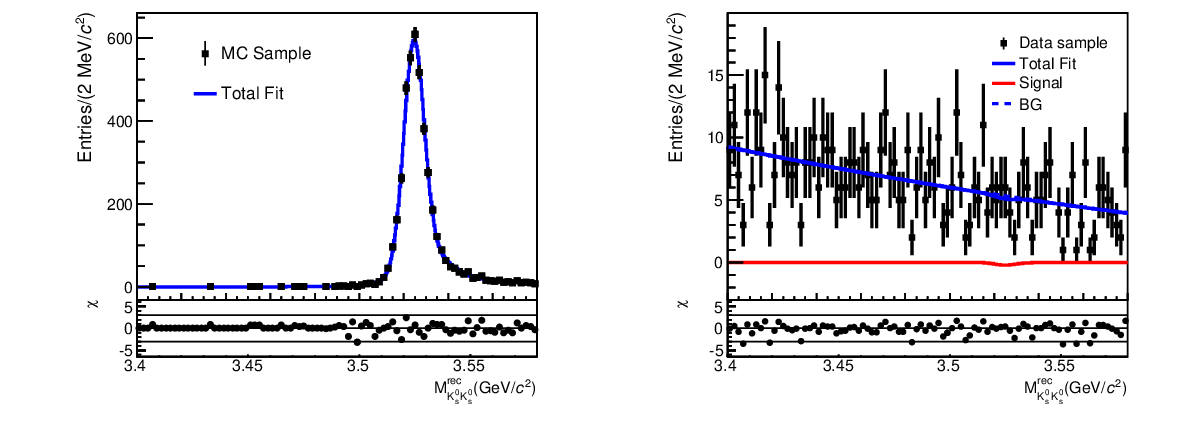}
\includegraphics[width=0.33\textwidth]{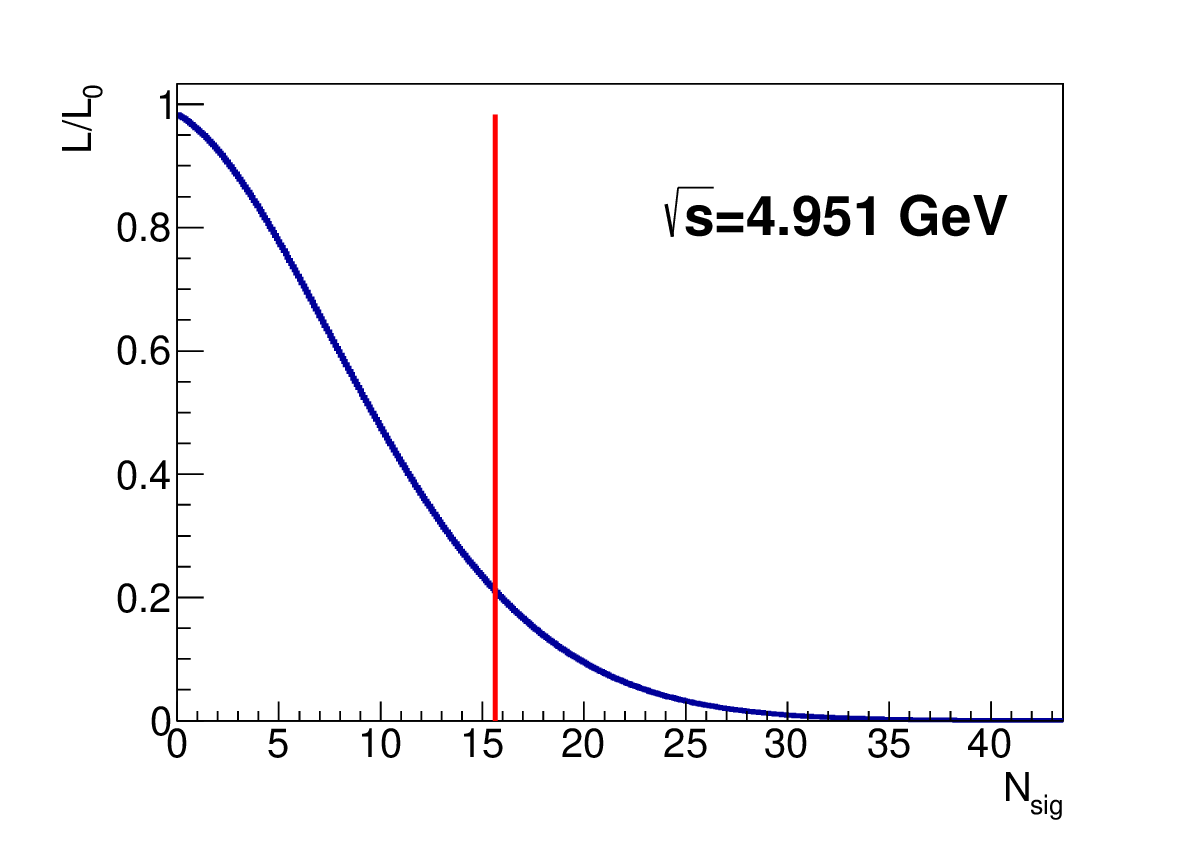}
\end{center}
\caption{Fit and scan results at $\sqrt{s}=4.843-4.951$~GeV.}
\label{fig:kskshc fit 4840}
\end{figure*}

\providecommand{\href}[2]{#2}\begingroup\raggedright

\newpage
{\bf \noindent The BESIII collaboration}\\

{\small
M.~Ablikim$^{1}$, M.~N.~Achasov$^{4,c}$, P.~Adlarson$^{76}$, O.~Afedulidis$^{3}$, X.~C.~Ai$^{81}$, R.~Aliberti$^{35}$, A.~Amoroso$^{75A,75C}$, Q.~An$^{72,58,a}$, Y.~Bai$^{57}$, O.~Bakina$^{36}$
% ...
\\
\vspace{0.2cm}
(BESIII Collaboration)\\
\vspace{0.2cm} 
{\it
$^{1}$ Institute of High Energy Physics, Beijing 100049, People's Republic of China\\
% ...
$^{3}$ Bochum  Ruhr-University, D-44780 Bochum, Germany\\
$^{4}$ Budker Institute of Nuclear Physics SB RAS (BINP), Novosibirsk 630090, Russia\\
% ...
$^{35}$ Johannes Gutenberg University of Mainz, Johann-Joachim-Becher-Weg 45, D-55099 Mainz, Germany\\
$^{36}$ Joint Institute for Nuclear Research, 141980 Dubna, Moscow region, Russia\\
% ...
$^{57}$ Southeast University, Nanjing 211100, People's Republic of China\\
$^{58}$ State Key Laboratory of Particle Detection and Electronics, Beijing 100049, Hefei 230026, People's Republic of China\\
% ...
$^{72}$ University of Science and Technology of China, Hefei 230026, People's Republic of China\\
% ...
$^{75}$ University of Turin and INFN, (A)University of Turin, I-10125, Turin, Italy; (B)University of Eastern Piedmont, I-15121, Alessandria, Italy; (C)INFN, I-10125, Turin, Italy\\
$^{76}$ Uppsala University, Box 516, SE-75120 Uppsala, Sweden\\
% ...
$^{81}$ Zhengzhou University, Zhengzhou 450001, People's Republic of China\\

\vspace{0.2cm}
$^{a}$ Deceased\\
% ...
$^{c}$ Also at the Novosibirsk State University, Novosibirsk, 630090, Russia\\
% ...

}

}
\end{document}